\documentclass[
  twocolumn,
  aps,
  prb,
  amsmath,
  amssymb,
  superscriptaddress,
  nofootinbib,
  floatfix
]{revtex4-1}
\usepackage{mathtools}
\usepackage{bm}
\usepackage{dsfont}
\usepackage{graphicx}
\usepackage{subfigure}
\usepackage{pifont}
\usepackage{textcomp}
\usepackage{gensymb}
\usepackage{accents}
\usepackage{upgreek}
\usepackage{array}
\usepackage{color}

\usepackage{hyperref}
\hypersetup{colorlinks=true, citecolor=blue, urlcolor=blue, linkcolor=blue}

\makeatletter
\newcommand*{\supplementarystart}{%
  \close@column@grid%
  \clearpage%
  \onecolumngrid%
  \setcounter{enumiv}{0} 
  \setcounter{equation}{0} 
  \setcounter{figure}{0} 
  \setcounter{table}{0} 
  \setcounter{page}{1}
  \c@secnumdepth=4
  \renewcommand{\theequation}{s\arabic{equation}} 
  \renewcommand{\bibnumfmt}[1]{[s##1]} 
  \renewcommand{\cite}[1]{{[}\onlinecite{##1}{]}}
  \renewcommand{\thefigure}{s\arabic{figure}}
  \renewcommand{\thetable}{s\Roman{table}}
  \renewcommand{\thepage}{s\arabic{page}}
}
\makeatother

\newcommand{\pa}{\partial}

\newcommand{\be}{\begin{equation}}
\newcommand{\e}{\end{equation}}
\newcommand{\beml}{\begin{subequations}}
\newcommand{\eml}{\end{subequations}}
\newcommand{\beq}{\begin{eqnarray}}
\newcommand{\eq}{\end{eqnarray}}
\newcommand{\ba}{\begin{array}}
\newcommand{\ea}{\end{array}}
\newcommand{\bpm}{\begin{pmatrix}}
\newcommand{\epm}{\end{pmatrix}}
\newcommand{\bc}{\begin{cases}}
\newcommand{\ec}{\end{cases}}
\newcommand{\lt}{\left}
\newcommand{\rt}{\right}
\newcommand{\n}{\nonumber}
\newcommand{\la}{\langle}
\newcommand{\ra}{\rangle}
\newcommand{\ep}{\varepsilon}

\newcommand{\bb}{\boldsymbol}
\newcommand{\h}{^\dagger}
\newcommand{\0}{^{\phantom{\dagger}}}

\DeclareMathOperator{\tr}{Tr}

\DeclareMathOperator{\im}{Im}

\begin{document}

\title{Spin-torque resonance due to diffusive dynamics at a surface of topological insulator}

\author{R.\,J.~Sokolewicz}
\affiliation{Radboud University, Institute for Molecules and Materials, NL-6525 AJ Nijmegen, the Netherlands}

\author{I.\,A.~Ado}
\affiliation{Radboud University, Institute for Molecules and Materials, NL-6525 AJ Nijmegen, the Netherlands}

\author{M.\,I.~Katsnelson}
\affiliation{Radboud University, Institute for Molecules and Materials, NL-6525 AJ Nijmegen, the Netherlands}
\affiliation{Theoretical Physics and Applied Mathematics Department, Ural Federal University, Mira Str. 19, 620002 Ekaterinburg, Russia}

\author{P.\,M.~Ostrovsky}
\affiliation{Max Planck Institute for Solid State Research, Heisenbergstr.\,1, 70569 Stuttgart, Germany}
\affiliation{L.\,D.~Landau Institute for Theoretical Physics RAS, 119334 Moscow, Russia}

\author{M.~Titov}
\affiliation{Radboud University, Institute for Molecules and Materials, NL-6525 AJ Nijmegen, the Netherlands}
\affiliation{ITMO University, Saint Petersburg 197101, Russia}

\begin{abstract}
We investigate spin-orbit torques on magnetization in an insulating ferromagnetic (FM) layer that is brought into a close proximity to a topological insulator (TI). In addition to the well-known field-like spin-orbit torque, we identify an anisotropic anti-damping-like spin-orbit torque that originates in a diffusive motion of conduction electrons. This diffusive torque is vanishing in the limit of zero momentum (i.\,e. for spatially homogeneous electric field or current), but may, nevertheless, have a strong impact on spin-torque resonance at finite frequency provided external field is neither parallel nor perpendicular to the TI surface. The required electric field configuration can be created by a grated top gate.
\end{abstract}

\maketitle

It is widely known that spin-orbit interaction provides an efficient way to couple electronic and magnetic degrees of freedom. It is, therefore, no wonder that the largest torque on magnetization, which is also referred to as the spin-orbit torque, emerges in magnetic systems with strong spin-orbit interaction \cite{miron_current-driven_2010,haney_current_2013} as has been long anticipated \cite{dyakonov_current-induced_1971}. 

The spin-orbit coupling may be enhanced by confinement potentials in effectively two-dimensional systems consisting of conducting and magnetic layers. The in-plane current may efficiently drive domain walls or switch magnetic orientation in such structures with the help of spin-orbit torque \cite{awschalom2009trend,manchon_theory_2008,garate_influence_2009,manchon2009theory}, which is present even for uniform magnetization, or with the help of spin-transfer torque, which requires the presence of magnetization gradient (due to e.\,g. domain wall) \cite{slonczewski_current-driven_1996,berger_emission_1996,ralph_spin_2008,stiles_anatomy_2002}. 

Topological insulators (TI) \cite{fu_topological_2007,moore_topological_2007,roy_topological_2009,hsieh_topological_2008} may be thought as materials with an ultimate spin-orbit coupling. Indeed, the effective Hamiltonian of conduction electrons at the TI surface contains essentially nothing but spin-orbit interaction term that provides a perfect spin-momentum locking. Thus, the magnetization dynamics in a thin ferromagnetic (FM) film in a proximity to TI surface is expected to be strongly affected by electric currents and/or electric fields \cite{qi_fractional_2008}. There seems to be, indeed, a substantial experimental evidence that the efficiency of domain switching in TI/FM heterostructures is dramatically enhanced as compared to that in metals \cite{mellnik_spin-transfer_2014,wang_SOT_BiSe_2014,fan_magnetization_2014,Fan_SOT_TI_2016,Yasuda_SOT_BiSbTe_2017,Cha2018}. 

Nowadays the symmetry of spin-orbit torques is routinely inferred from the ferromagnetic resonance measurements in which an alternating microwave-frequency current (with frequencies $7-12$\,GHz) is applied within the sample plane \cite{mellnik_spin-transfer_2014,Ralph2011SOTFMresonance,Wang2015_SOTHflCoFEBIMgO,Ralph2016SOTWTE2,Ralph2018SOT}. 

In this work we identify a novel anti-damping-like torque originating in a diffusive motion of conduction electrons at the TI surface. Such a torque originates in a non-local diffusive response of $z$ component of the conduction electron spin density to the in-plane electric field. The non-locality of the response is determined by the so-called diffusion pole in analogy to the density-density response of a disordered system. It is, however, important that the diffusive response of the spin-density in the TI is always present in the perpendicular-to-the-plane component of the spin density, irrespective of the magnetization direction in the FM. In non-topological FM/metal systems such a diffusive response is present only in the spin density component that is directed along the local magnetization of the FM. Thus, the diffusive anti-damping spin-orbit torque, that we describe below, is specific for the TI/FM interfaces. Similarly, we identify a strong anisotropy of the Gilbert damping in the TI/FM system due to a combination of electron elastic scattering on non-magnetic impurities and a spin-momentum locking in the TI.

\begin{figure}
\centering
\centerline{\hspace*{1cm}\includegraphics[width=0.8\linewidth]{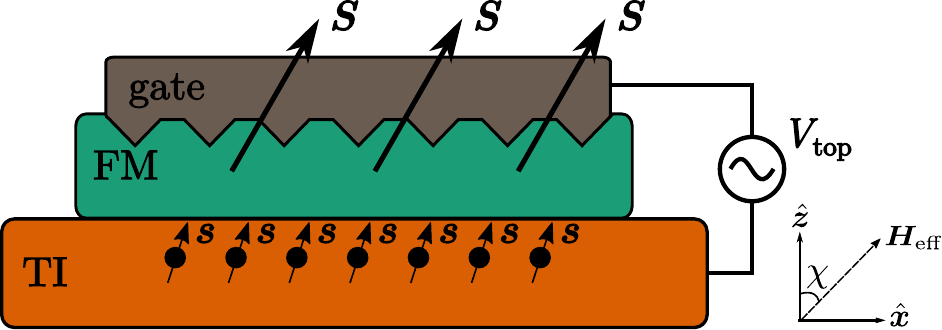}}
\caption{Proposed experimental setup. Non-homogeneous in-plane electric field components are created by an $ac$ top-gate voltage $V_{\rm{top}}$ that  induce a strong diffusive spin-orbit torque (\ref{diffSOT}) of the damping-like symmetry. An effective magnetic field $\bb{H}$ is directed at the angle $\chi$ with respect to $\hat{\bb{z}}$.}
\label{fig:setup}
\end{figure}

Diffusive anti-damping spin-orbit torque, that we are going to study, can be related to a response of conduction electron spin density to electric field at a finite, but small, frequency and momentum. Such a field can be created e.\,g. by applying an \textit{ac} gate voltage to a grated top-gate as shown in Fig.~\ref{fig:setup}. The presence of the diffusive spin-orbit torque can be detected by rather unusual spin-orbit-torque resonances in the TI/FM structures that we also investigate in this work. 

Microscopic theory of current-induced magnetization dynamics in TI/FM heterostructures has been so far limited to some particular cases: (i) specific direction of magnetization and (ii) the limit of vanishing exchange interaction between FM angular momenta and the spins of conduction electrons.  In particular, an analytic estimate of spin-transfer and spin-orbit torques in TI/FM bilayer has been given in Ref.~\cite{sakai_spin_2014} for magnetization perpendicular to the TI surface. An attempt to generalize these results to arbitrary magnetization direction has been undertaken more recently in Ref.~\cite{ndiaye_dirac_2017}.  The non-local transport on a surface of the TI has been first discussed in Ref.~\cite{burkov_spin_2010}. The results of this work has been later applied to TI/FM systems \cite{taguchi_spin-charge_2015,shintani_spin_2016} in a perturbative approach with respect to a weak s--d-type exchange. The non-local behavior of non-equilibrium out-of-plane spin polarization in TI/FM systems, which gives rise to diffusive spin-orbit torques, have been, however, overlooked in all these publications. 

To describe magnetization dynamics at a TI/FM interface we employ an effective two-dimensional Dirac model for conduction electrons 
\be
\label{TImodel}
\mathcal{H}= v\,\lt[(\bb{p}-e\bb{A})\times \bb{\sigma}\rt]_z - \Delta_\textrm{sd}\, \bb{m}\cdot \bb{\sigma} +V(\bb{r}),
\e
where $\bb{A}$ stands for the vector potential, $e=-|e|$ is the electron charge, $z$ is the direction perpendicular to the TI surface, $v$ is the effective velocity of Dirac electrons, and $V(\bb{r})$ is a disorder potential that models the main relaxation mechanism of conduction electrons. The energy $\Delta_{\textrm{sd}}=J_\textrm{sd} S$ is characterizing the local exchange interaction $\mathcal{H}_\textrm{ex}=-J_\textrm{sd} \sum_n  \bb{S}_n \cdot c\h_n \bb{\sigma}c\0_n$ between localized classical magnetic moments $\bb{S}_n$ on FM lattice (with conserved absolute value $S=|\bb{S}_n|$ per unit cell area $\mathcal{A}$) and the electron spin density  (represented by the vector operator $\bb{\sigma} = (\sigma_x,\sigma_y,\sigma_z)$ on the TI surface) \cite{sdmodel}. Here $\sigma_\alpha$ stand for Pauli matrices and $J_\textrm{sd}$ quantifies the s--d-type exchange interaction strength.  

Classical equation of motion for the unit magnetization vector $\bb{m}=\bb{S}/S$ is determined by the s-d like exchange interaction $\mathcal{H}_\textrm{ex}$ as
\be
\label{Eom}
\pa \bb{m}/\pa t = \gamma\, \bb{H} \times\bb{m}+\bb{T},\quad
\bb{T}=(J_\textrm{sd}\mathcal{A}/\hbar) \,\bb{m}\times\bb{s},
\e
where $\hbar=h/2\pi$ is the Planck constant and $\gamma$ is a gyromagnetic ratio for the FM spin. The effective field $\bb{H}$ represents the combined contribution of external magnetic field and the field produced by neighboring magnetic moments in the FM (e.\,g. due to direct exchange), while the term $\bb{T}$ represents the effect of the conduction electron spin density $\bb{s}(\bb{r},t)= \la c\h_n \bb{\sigma} c\0_n\ra$  on the TI surface. 

To quantify the leading contributions to $\bb{T}$ we microscopically compute: i) a linear response of $\bb{s}$ to the in-plane electric field $\bb{E}(\bb{r},t)= \bb{E}_{\bb{q},\omega}\exp\lt(-i\omega t+i\bb{q}\cdot\bb{r}\rt)$; and ii) a linear response of $\bb{s}$  to the time derivative $\pa \bb{m}/\pa t$. The former response defines the spin-orbit torque, while the latter one does the Gilbert damping. 
 
Before we proceed with the analysis we shall note that the velocity operator $\bb{v} = v\,(\bb{\sigma}\times \hat{\bb{z}})$ in the model of Eq.~(\ref{TImodel})  is directly related to the spin operator $\bb{\sigma}$. As the result, the response of the in-plane spin density $\bb{s}_\parallel=(s_{x}, s_{y})$ to electric field $\bb{E}=-\pa\bb{A}/\pa t$ is defined by the conductivity tensor \cite{ndiaye_dirac_2017,Ghosh18}. This also means that the non-equilibrium contribution to $\bb{s}_\parallel$ from the electric current density $\bb{J}$ is given by $\bb{s}_\parallel= (\hat{\bb{z}}\times\bb{J} )/ev$ for any frequency and momentum irrespective of type of scattering for conduction electrons and even beyond the linear response. 

Thus, the response of $\bb{s}_\parallel$ defines an exceptionally universal field-like spin-orbit torque
\be
\label{TFL}
\bb{T}^\textrm{SOT}_\textrm{FL}= (J_\textrm{sd}\mathcal{A}/\hbar e v)\, \bb{m}\times(\hat{\bb{z}}\times\bb{J}),
\e 
that acts in the same way as in-plane external magnetic field applied perpendicular to the charge current. 

Apart from the universal response of $\bb{s}_\parallel$ there might also exists a non-equilibrium spin polarization $s_z$ perpendicular to the TI surface. This component plays no role in Eq.~(\ref{Eom}) for $\bb{m}= \pm \hat{\bb{z}}$ due to the vector product involved. Also, the $s_z$ component is vanishing by symmetry for $\bb{m}=\bb{m}_\parallel$, where we decompose $\bb{m} =\bb{m}_\parallel+\bb{m}_\perp$ to in-plane and perpendicular-to-the plane components.

We find, however, that for a general direction of $\bb{m}$, the spin density $s_z$ may be strongly affected by the in-plane electric field at a small but finite frequency and a small but finite wave vector. In the leading approximation the result can be cast in the following form 
\be
\label{diffSOT}
\bb{T}^\textrm{SOT}_\textrm{diff} = \eta\, \bb{m}\times\bb{m}_\perp\, \frac{i D\, \bb{q}\cdot \bb{E}}{i\omega -Dq^2},\quad
\eta=\frac{eJ_\textrm{sd}^2\mathcal{A}S}{2\pi \hbar^3v^2},
\e
where $D$ is a diffusion coefficient for conduction electrons at the TI surface and $\bb{E}= \bb{E}_{\bb{q},\omega}\exp\lt(-i\omega t+i\bb{q}\cdot\bb{r}\rt)$. Note that the diffusive torque is non-linear with respect to $\bb{m}$ and, from the point of view of the time reversal symmetry, is analogous to anti-damping torque. The denominator $i\omega -Dq^2$ in Eq.~(\ref{diffSOT}) reflects diffusive (Brownian) motion of conduction electrons that defines the time-delayed diffusive torque on magnetization $\bb{T}^\textrm{SOT}_\textrm{diff}$.  

It is interesting to note that the torque of Eq.~(\ref{diffSOT}) has an anti-damping symmetry (when expressed through electric current rather than electric field). Moreover, the torque formally diverges as $1/q$ in the $dc$ limit $\omega=0$. This singularity is well-known in the theory of disordered systems \cite{maleev1975corrections,m*aleev1976corrections,altshuler1985electron} and originates in the diffusive (Brownian) motion of conduction electrons in a disorder potential.  The $dc$ limit singularity in Eq.~(\ref{diffSOT}) is, in fact, regularized by the dephasing length of conduction electrons on the surface of the TI. The length is strongly temperature and material dependent and, at low temperatures, can reach hundreds of microns. Thus, the result of Eq.~(\ref{diffSOT}) also predicts large anti-damping spin-orbit torque in the dc-limit that originates in a mechanism which is specific for the TI interface. 
 
In order to derive the result of Eq.~(\ref{diffSOT}) and the expressions for Gilbert damping we shall adopt a particular relaxation model for both spin and orbital angular momenta of conduction electrons. For the model of Eq.~(\ref{TImodel}) those are provided by scattering on disorder potential. We choose the latter 
to be the white-noise Gaussian disorder potential that is fully characterized by a single dimensionless parameter $\alpha\ll 1$,
\be
\label{disorder}
\la V(\bb{r})\ra =0,\quad \la V(\bb{r}) V(\bb{r}') \ra = 2\pi \alpha\, (\hbar v)^2\,\delta(\bb{r}-\bb{r}'),
\e
where angular brackets stay for the averaging over the ensemble of disordered systems. 

Since both the vector potential $\bb{A}$ and the magnetization $\bb{m}$ couple to spin operators in Eq.~(\ref{TImodel}), the linear response of $\bb{s}$ to $\bb{E}=-\pa\bb{A}/{\pa t}$ and $\pa \bb{m}/\pa t$ is defined in the frequency-momentum domain as 
\be
\label{spol}
\bb{s} =(v^2 h)^{-1} \hat{K}(\bb{q},\omega) \lt[e v\,(\bb{E}\times \hat{\bb{z}}) - i \omega\Delta_\textrm{sd}\,\bb{m} \rt].
\e
Here, the dimensionless 9-component tensor $\hat{K}(\bb{q},\omega)$ is given by the Kubo formula 
\be
\label{Kubo}
\hat{K}_{\alpha\beta}(\bb{q},\omega)= {v^2}\!\!\int\!\! \frac{d^2\bb{p}}{(2\pi)^2}
\tr \lt\la \sigma_\alpha G^\textrm{R}_{\bb{p}+\hbar \bb{q},{\ep+\hbar\omega}}\sigma_\beta G^\textrm{A}_{\bb{p},\ep}\rt \ra,
\e
where the notation $G^{\textrm{R(A)}}_{\bb{p},\ep}$ stands for the retarded (advanced) Green's function for the Hamiltonian of Eq.~(\ref{TImodel}), the angular brackets denote the averaging over disorder realizations, while the energy $\ep$ refers to the Fermi energy (zero temperature limit is assumed). 

The tensor $\hat{K}$ can be represented by the matrix 
\be
\label{Kten}
\hat{K}=  
\bpm \sigma_{xx} & \sigma_{xy} & Q_y \\ \sigma_{yx} & \sigma_{yy} & -Q_x \\ Q_y & -Q_x & \zeta \epm,
\e
of which $\sigma_{\alpha\beta}$ are the components of the two dimensional conductivity tensor at the TI surface (all conductivities are expressed in the units of $e^2/h$),  the vector $\bb{Q}$ defines the diffusive spin-orbit torque of Eq.~(\ref{diffSOT}) (its contribution to Gilbert damping is negligible), while $\zeta$ determines the response of $s_z$ to $\pa m_z/\pa t$. The components of $\hat{K}$ correspond to different responses at different limits. When discussing the response to an electric field $\bb{E}_{\bb{q}\omega}$ we are primarily interested in the limit $\omega\ll Dq^2$, whereas the response to time-derivative of magnetization $\bb{m}$ is defined by the limit $q\to 0$.

In the linear response theory of Eq.~(\ref{spol}) one needs to compute the tensor in Eq.~(\ref{Kubo}) for a constant direction $\bb{m}$ and for $\bb{A}=0$. In usual systems (conducting ferromagnets) the response of $\bb{s}$ in the direction of $\bb{m}$ is always diffusive. This response, however, plays no role in the torque since $\bb{T}\propto \bb{m}\times \bb{s}$. The situation at the TI surface is, however, special. Here, the in-plane components of magnetization $m_x, m_y$ play no role in Eq.~(\ref{TImodel}), since those are simply equivalent to a constant in-plane vector potential for conduction electrons and, therefore, can be excluded by a gauge transform (shift of the Dirac cone). Consequently, all observable quantities in the model (including all components of the tensor $\hat{K}$) may only depend on the field $\Delta_z=\Delta_\textrm{sd} m_z$. As the result, the diffusive response occurs exclusively in $s_z$ component of spin polarization and can easily enter the expression for the torque.

The conductivity tensor in the model of Eqs.~(\ref{TImodel},\ref{disorder}) has been analyzed in detail in Ref.~\cite{ivanEPL} in the limit $\omega=\bb{q}=0$ (and for $\alpha\ll 1$) with the result $\sigma_{xx}=\sigma_{yy}=\sigma_0$ and $\sigma_{xy}=-\sigma_{yx}=\sigma_\textrm{H}$, where  
\be
\label{cond}
\sigma_0= \frac{\ep^2-\Delta_z^2}{\pi\alpha\,(\ep^2+3\Delta_z^2)},\qquad \sigma_\textrm{H}= \frac{8 \ep \Delta_z^3}{(\ep^2+3\Delta_z^2)^2}.
\e
Since the anomalous Hall conductivity $\sigma_\textrm{H} \propto \alpha\, \sigma_0$ is sub-leading with respect to $\sigma_0$, it has to be computed beyond the Born approximation (see Refs.~\cite{ivanEPL,ivanPRL,ivanPRB}). 

Here we generalize the analysis to calculate the tensor $\hat{K}$ for finite $\omega$ and $\bb{q}$ assuming $\alpha\ll 1$, $\omega\tau_\textrm{tr}\ll 1$, and $\omega \propto Dq^2$, where $D=\hbar v^2  \sigma_0/\ep$ is the diffusion coefficient and $\tau_\textrm{tr}=\hbar\ep\sigma_0/(\ep^2+\Delta_z^2)$ is the transport scattering time for the problem. In real samples $\tau_\textrm{tr}=0.01 - 1$\,ps \cite{kong_rapid_2011,kamboj_probing_2017,huang_enhancement_2017,xiang_transport_2014}.

The main building block of our analysis is the averaged Green's function in the first Born approximation
\be
\label{green}
G^{\mathrm{R}}_{\bb{p},\ep} = \frac{\ep^\textrm{R} + v(\bb{p}\times \bb{\sigma})_z-\Delta^{\textrm{R}}_z\sigma_z}
{(\ep^{\textrm{R}})^2-v^2p^2 -(\Delta^{\textrm{R}}_z)^2},
\e
where the complex parameters $\ep^\textrm{R}=\ep(1+i\pi \alpha/2)$ and  $\Delta^\textrm{R}_z= \Delta_z(1-i\pi \alpha/2)$ are found from the corresponding self-energy 
\be
\label{Sigma}
\Sigma^{\mathrm{R}}(\ep)  = 2\pi \alpha\,v^2\!\! \int\!\frac{d^2\bb{p}}{(2\pi)^2}G^\mathrm{R}_{\bb{p},\ep}, 
\e
that gives rise to  $\im \Sigma^{\mathrm{R}}= - \pi \alpha (\ep -\Delta_z \sigma_z)/2$  (strictly speaking, the RG analysis \cite{ivanEPL} has to be applied). In the Green's function of Eq.~(\ref{green}) we shift the momentum $\bb{p}$ such that there is no direct dependence on the in-plane magnetization components $m_x$ and $m_y$. 

\begin{figure}[t]
\centering
\includegraphics[]{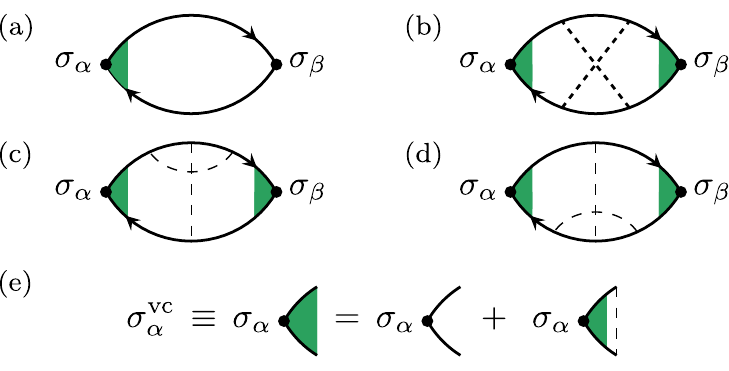}
\caption{Diagrams considered in the calculation of $\hat{K}$: (a) non-crossing diagram, (b) $X$ diagram, (c-d) $\Psi$ diagrams. Green areas indicate the ladder summation (e) for the vertex correction in the non-crossing approximation \cite{ivanEPL}.}
\label{fig:diagrams}
\end{figure}

The next step in disorder-averaging requires the computation of vertex corrections. This means we need to replace the spin operator $\sigma_\alpha$ with a vertex corrected spin operator $\sigma_\alpha^\text{vc}$ in the ladder approximation as depicted in Fig.~\ref{fig:diagrams}(e). The crossed diagrams in Fig.~\ref{fig:diagrams}(b-d) give a contribution to the components of $\hat{K}$ of the order $\mathcal{O}(\alpha^0)$. The
only components that are modified to this order are those
corresponding to the Hall conductivity (i.e. $\sigma_{xy}$ and $\sigma_{yx}$). Details of this calculation can be found Ref.~\cite{ivanEPL}.

The dressing of $\sigma_\alpha$ with a single disorder line is denoted by $\sigma_\alpha^{1\times\text{dr}}$ and is conveniently represented in the matrix form by introducing a matrix $\hat{M}$ with $16$ components $M_{\alpha\beta}$ for $\alpha,\beta=0,x,y,z$ (with $\sigma_0=1$)
    \begin{align}
       \sigma_\alpha^{1\times\text{dr}}  = 2\pi \alpha\, v^2 \int\!\frac{d^2\bb{p}}{(2\pi)^2} G^\text{A}_{\ep+\omega,\bb{p}+\bb{q}}\sigma_\alpha G^\text{R}_{\bb{p}} = \pi\alpha M_{\alpha\beta}\sigma_\beta,
        \label{eq:myseries}
    \end{align}
where the summation of the repeating index $\beta=0,x,y,z$ is assumed. Full expressions of the components of $\hat{M}$ up to second order in $\omega$ and $q$ are given by Eq.~(\ref{eq:M}a-f). 

In our calculation the terms of the order of $\alpha \ln p_\textrm{cutoff}/\ep$ (where $p_\textrm{cutoff}$ is the ultraviolet momentum cut-off) is disregarded with respect to $1$. This approximation is legitimate since we assume that all model parameters $\epsilon$, $\Delta_\textrm{sd}$ and $\alpha$ are first renormalized such that $p_\textrm{cutoff} \approx \ep$.

It is, then, easy to see that the vertex-corrected spin operator is readily obtained from the geometric series of powers of $\pi\alpha \hat{M}$, 
\begin{align}
\sigma_\alpha^\text{vc} &= 
\sigma_\alpha+\pi\alpha \hat{M}_{\alpha\beta}\sigma_\beta+(\pi\alpha)^2 (\hat{M}^2)_{\alpha\beta}\sigma_\beta+\dots\nonumber\\
&=\left[1-\pi\alpha \hat{M}\right]^{-1}_{\alpha\beta}\sigma_\beta.
\end{align}
Thus, in the non-crossing approximation (illustrated in Fig.~\ref{fig:diagrams} (a)), one simply finds $\hat{K}= \hat{M}[1-\pi\alpha \hat{M}]^{-1}$. 

Dressed spin-spin correlators are defined by the components $ \hat{K}_{\alpha\beta}$ with $\alpha,\beta=x,y,z$.
The vector $\bb{q}$ selects a particular direction in space, that makes the conductivity tensor anisotropic. By choosing $x$ direction along the $\bb{q}$ vector, we find the conductivity components $\sigma_{xx} = \sigma_0$, $\sigma_{xy} = -\sigma_{yx}=\sigma_\textrm{H}$, and $\sigma_{yy}= i\omega \,\sigma_0/(i\omega-Dq^2)$, where we have kept only the leading terms in the limits $\alpha\ll 1$, $\omega\tau_\textrm{tr}\ll 1$ (more general expressions are given by Eqs.~(\ref{eq:sconductivity}a-d)). We can see that $\sigma_{yy}$ component also acquires a diffusion pole.  One needs to go beyond the non-crossing approximation in the computation of anomalous Hall conductivity  \cite{ivanEPL,ivanPRL,ivanPRB}. 

Clearly, the components $\sigma_{\alpha\beta}$ define the field-like contribution $\bb{T}^\textrm{SOT}_\textrm{FL}$ that has been already discussed above. It is interesting to note that the conductivity is isotropic $\sigma_{xx}=\sigma_{yy}=\sigma_0$ only if the limit $q= 0$ is taken before the limit $\omega=0$. If the limit $\omega=0$ is taken first, the conductivity remains anisotropic with respect to the direction of $\bb{q}$ even for $q=0$.

The vector $\bb{Q}=(Q_x,Q_y)$ quantifies both the response of $s_z$ to electric field or to $\pa \bb{m}_\parallel/\pa t$ as well as the response of $\bb{s}_\parallel$ to  $\pa m_z/\pa t$. From Eq.~(\ref{Kubo}) we find,
\be
\label{Qvec}
\bb{Q}(\omega,\bb{q})=\frac{\Delta_z}{\hbar v}\, \frac{i D \bb{q}}{i\omega-Dq^2}\lt(1+\mathcal{O}(\omega\tau_\textrm{tr})\rt),
\e
where we again assumed $\omega\tau_\textrm{tr} \ll 1$. The result of Eq.~(\ref{Qvec}), then, corresponds to an additional diffusive spin-orbit torque of the form Eq.~(\ref{diffSOT}).

Finally, the response of $s_z$ to $\pa m_z/\pa t$ is defined by 
\be
\label{szz}
\zeta = \frac{\Delta_z^2}{i\hbar\ep\omega}\lt(1+\mathcal{O}(\omega^2\tau^2_\textrm{tr})\rt),
\e
where the limit $q=0$ is taken. Thus, we find from Eq.~(\ref{spol}) that there exists no response of $s_z$ to $\pa m_z/\pa t$. Instead, the quantity $\zeta$ defines the additional spin polarization in $z$-direction $\delta s_z = - \Delta_\textrm{sd}^3m_z^3/(2\pi \hbar^2 v^2\ep)$ that we ignore below. Eqs.~(\ref{Qvec},\ref{szz}) including subleading terms in $\omega\tau_\mathrm{tr}$ are presented in Eq.~(\ref{Qresult}).

We also note, that $\bb{Q}(q=0)=0$, hence there is no term in $s_z$ that is proportional to $\pa\bb{m}/\pa t$. This reflects highly anisotropic nature of the Gilbert damping in the model of Eq.~(\ref{TImodel}).

The remaining parts of the Gilbert damping can be cast in the following form 
\be
\bb{T}^\textrm{GD}=\frac{J^2_\textrm{sd} \mathcal{A} S}{\pi\hbar^2 v^2}\,\bb{m}\times
\Big(\sigma_0\,\frac{\pa \bb{m}_\parallel}{\pa t} + \frac{\sigma_\textrm{H}}{m_z}\, \frac{\pa \bb{m}_\parallel}{\pa t}\times\bb{m}_\perp \Big),
\label{GD}
\e
where the coefficients, $\sigma_0$ and $\sigma_\textrm{H}/m_z$ from Eq.~(\ref{cond}) depend on $m_z^2$, which is yet another source of the Gilbert damping anisotropy.  We note, that even though Eq.~(\ref{GD}) does not contain a term proportional to $\partial m_z/\partial t$, the existing in-plane Gilbert damping is sufficient to relax the magnetization along $\hat{\bb{z}}$ direction. 

Despite strongly anisotropic nature of the diffusive torque (the torque is vanishing for purely in-plane or purely perpendicular to the plane magnetization), its strength for a generic direction of magnetization may be quite large. For example, for $\bb{m}$ directed approximately at 45 degrees to the TI surface the ratio of amplitudes of diffusive and field like torques is readily estimated as  
\be
\label{estimate}
\frac{T^\textrm{SOT}_\textrm{diff}}{T^\textrm{SOT}_\textrm{FL}} = \frac{\Delta_\textrm{sd}}{\hbar q v}\,\frac{1}{\sigma_0},
\e
where we used the condition $\omega \ll Dq^2$.  Let us assume that a top-gate in Fig.~\ref{fig:setup} induces an ac in-plane electric field with the characteristic period $2\pi q^{-1}\approx 1$\,$\mu$m and a typical FM resonance frequency, $\omega\approx 7$-$12$\,GHz. Then, for realistic materials one can estimate $Dq^2\approx 100$\,GHz, hence $\omega \ll Dq^2$ indeed. For a typical velocity $v=10^6$\,m$/$s one finds  $\hbar q v\approx 4$\,meV. Thus, the ratio $\Delta_\textrm{sd}/\hbar q v$ in Eq.~(\ref{estimate}) may reach three orders of magnitude, while the value of $\sigma_0$ is typically $10$. This estimate suggests that, for a generic direction of $\bb{m}$, the magnitude of diffusive  torque can become three orders of magnitude larger than that of the field-like spin-orbit torque. 

The diffusive torque at the TI surface can be most directly probed by the corresponding spin-torque resonance. In this case, one can disregard the effect of the field like torque, so that Eq.~(\ref{Eom}) is simplified to
\be
\label{equation}
\frac{\pa \bb{m}}{\pa t}=\gamma\,\bb{H}\times \bb{m}+f(\bb{r},t)\,\bb{m}\times\bb{m}_\perp+\alpha_\textrm{G}\,\bb{m}\times\frac{\pa \bb{m}_\parallel}{\pa t},
\e
where $\alpha_\textrm{G}= J_\textrm{sd}^2 \mathcal{A}S \sigma_0/\pi(\hbar v)^2$ is the Gilbert damping amplitude (which is a constant for $\ep \gg \Delta_\textrm{sd}$), while the terms containing $\sigma_\textrm{H}$ are omitted.  The function
\be
f(\bb{r},t)=\eta\int\! d^2\bb{r}'\!\!\int_{-\infty}^t\!\!\!\! dt'\, \frac{e^{-(\bb{r}-\bb{r}')^2/4D(t-t')}}{4\pi(t-t')}\bb{\nabla}\cdot\bb{E}(\bb{r}',t'),\n
\e
defines the strength of the diffusive spin-orbit torque (\ref{diffSOT}) in real space and time.

\begin{figure}
\centering
\centerline{\hspace*{1cm}\includegraphics[width=1.0\linewidth]{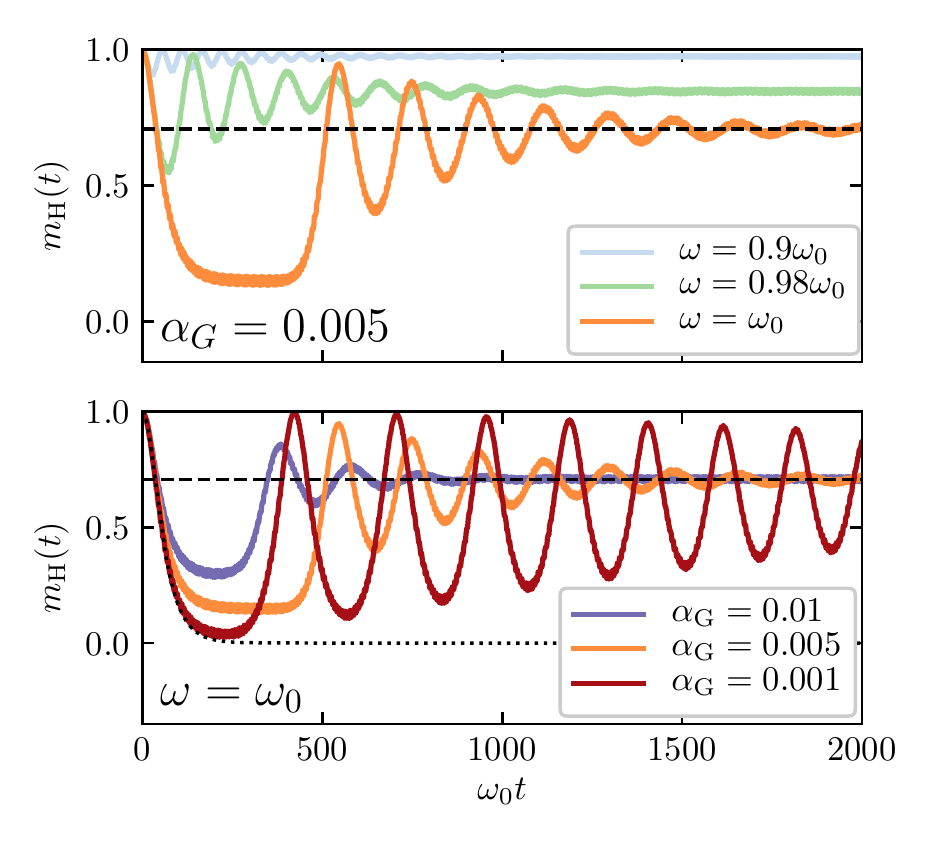}}
\caption{The projection $m_\textrm{H}(t)$ as simulated from Eq.~(\ref{equation}) for $f_0=0.1\, \omega_0$.  Top panel illustrates the behavior at different frequencies for $\alpha_\textrm{G}=0.005$. Lower panel illustrates the resonant behavior at different values of $\alpha_\textrm{G}$. Dashed horizontal line corresponds to $m_\textrm{H}=1/\sqrt{2}$. Dots indicate the asymptotic solution for $\alpha_\textrm{G}=0$ as given by Eq.~(\ref{asymptot}).}
\label{fig:dynamics}
\end{figure}

Resonant magnetization dynamics defined by Eq.~(\ref{equation}) is illustrated in Fig.~\ref{fig:dynamics} for $\bb{H}$ directed at the angle $\chi =\pi/4$ with respect to $\hat{\bb{z}}$ and for frequencies that are close to the resonant frequency $\omega_0=\gamma H$. The time evolution of magnetization projection $m_\textrm{H}= \bb{m}\cdot \bb{H}/H$ is induced by the diffusive torque with $f(t)= f_0\,\cos\omega t$ (magnetization at different $\bb{r}$ is simply different by a phase). 

Resonant dynamics at $\omega=\omega_0$ in Eq.~(\ref{equation}) consists of precession of $\bb{m}$ around the vector $\bb{H}$ such that the azimut (precession) angle is changing linearly with time $\phi(t)=\omega_0 t-\pi/2$  (for $f_0\ll \omega_0$ and $\alpha_\textrm{G}\ll1$). In addition, the projection $m_\textrm{H}$ oscillates between $1$ and $0$ on much larger time scales. Such oscillations are damped by a finite $\alpha_\textrm{G}$ to the limiting value $m_\textrm{H}=1/\sqrt{2}$. 

In the limit of vanishing Gilbert damping, $\alpha_\textrm{G}=0$, one simply finds the result 
\be
\label{asymptot}
m_\textrm{H}(t)=\lt[\cosh\lt(\tfrac{1}{4}f_0 t \sin (2\chi) \rt)\rt]^{-1},
\e
which clearly illustrates the absence of the effect for both perpendicular-to-the-plane ($\chi=0$) and in-plane ($\chi=\pi/2$) magnetization. The qualitative behavior at the resonance ($\omega=\omega_0$) is illustrated at the lower panel of Fig.~\ref{fig:dynamics} for different values of $\alpha_\textrm{G}$.

In conclusion, we consider magnetization dynamics in a model TI/FM system at a finite frequency $\omega$ and $\bb{q}$ vector. We identify novel diffusive anti-damping spin-orbit torque that is specific to TI/FM system.  Such a torque is absent in usual (non-topological) FM/metal systems, where the diffusive response of conduction electron spin density is always aligned with the magnetization direction of the FM. In contrast, the electrons at the TI surface gives rise to singular diffusive response of the conduction electron spin-density in the direction perpendicular to the TI surface, irrespective of the FM magnetization direction. Such a response leads to strong non-adiabatic anti-damping spin-orbit torque that has a diffusive nature. This response is specific for a system with an ultimate spin-momentum locking and gives rise to abnormal anti-damping diffusive torque that can be detected by performing spin-torque resonance measurements. We also show that, in realistic conditions, the anti-damping like diffusive torque may become orders of magnitude larger than the usual field like spin-orbit torque. We investigate the peculiar magnetizations dynamics induced by the diffusive torque at the frequency of the ferromagnet resonance. Our theory also predicts ultimate anisotropy of the Gilbert damping in the TI/FM system. In contrast, to the phenomenological approaches \cite{vanderBijl2012,Hals2013} our microscopic theory is formulated in terms of very few effective parameters. Our results are complementary to previous phenomenological studies of Dirac ferromagnets \cite{tserkovnyak_theory_2009,mahfouzi_spin-orbit_2012,katsnelson15,fischer_spin-torque_2016,yokoyama_theoretical_2010,yokoyama_current-induced_2011,siu_spin_2016,mahfouzi_antidamping_2016,soleimani_spin-orbit_2017,kurebayashi_microscopic_2017,chen_current-induced_2017,rodriguez-vega_giant_2016,qi_topological_2008,garate_inverse_2010,yokoyama_theoretical_2010,yokoyama_current-induced_2011,nomura_electric_2010,tserkovnyak_thin-film_2012-1,linder_improved_2014,tserkovnyak_spin_2015,ueda_topological_2012,liu_reading_2013,chang_nonequilibrium_2015,fischer_spin-torque_2016,mahfouzi_antidamping_2016,fujimoto_transport_2014,okuma_unconventional_2016}.

\begin{acknowledgments}
We are grateful to O.~Gomonay, R.~Duine, and J.~Sinova for helpful discussions. This research was supported by the JTC-FLAGERA Project GRANSPORT and by the Dutch Science Foundation NWO/FOM 13PR3118. M.T. acknowledges the support from the Russian Science Foundation under Project 17-12-01359. 

\end{acknowledgments}

%
%


%
%

\appendix
\section{Kubo formula}
The linear response formula used in the main text can be obtained in a Keldysh-framework. We start by introducing the Green function $\mathcal{G}$ in rotated Keldysh space [see e.g. Ref.~\cite{rammer_quantum_1986}]
\begin{equation}
    \mathcal{G}=\begin{pmatrix}G^\text{R}&G^\text{K}\\0&G^\text{A}\end{pmatrix}
\end{equation}
where R, A and K denote retarded, advanced and Keldysh Green functions respectively. In this notation a perturbation to a classical field $V(\bb{x},t)$ is given by
\begin{multline}
    \delta\mathcal{G}(\bb{x}_1,t_1;\bb{x}_2,t_2)=\int\!d\bb{x}_3\int\!dt_3\, \mathcal{G}^{(0)}(\bb{x}_1,t_1;\bb{x}_3,t_3)\\\hat{V}(\bb{x}_3,t_3)\mathcal{G}^{(0)}(\bb{x}_3,t_3;\bb{x}_2,t_2)+\mathcal{O}(V^2)
    \label{eq:s2}
\end{multline}
with $\mathcal{G}^{(0)}$ equilibrium Green functions. The Wigner-transform of a function $F(\bb{x}_1,t_1;\bb{x}_2,t_2)$ is given by
  \begin{multline}
    F(\bb{x}_1,t_1;\bb{x}_2,t_2)=\int\frac{d^2\bb{p}}{(2\pi\hbar)^2}\int \frac{d\ep}{2\pi\hbar}\,e^{-i\ep(t_1-t_2)/\hbar}\\e^{i\bb{p}\cdot(\bb{x}_1-\bb{x}_2)/\hbar}F(\ep,\bb{p},R,T)
   \end{multline}
with energy $\ep$, momentum $\bb{p}$, time $T=\frac{t_1+t_2}{2}$ and position $\bb{R}=\frac{\bb{x}_1+\bb{x}_2}{2}$. In equilibrium the Green functions $\mathcal{G}^{(0)}$ do not depend on $\bb{R}$ and $T$, so that the momentum-frequency representation of Eq.~(\ref{eq:s2}) becomes 
   $\delta\mathcal{G}(\ep,\omega,\bb{p},\bb{q})= \mathcal{G}^{(0)}_{\ep_+,\bb{p}_+} V_{\omega,\bb{q}}\mathcal{G}^{{(0)}}_{\ep_-,\bb{q}_-}$, 
with subscripts $\ep_\pm=\ep\pm\hbar\omega/2$ and $\bb{p}_\pm=\bb{p}\pm\hbar\bb{q}/2$ and $V_{\omega,\bb{q}}$ the Fourier transform of $V(\bb{R},T)$. 

The spin density $\bb{s}_{\omega,\bb{q}}$ is given by
\begin{equation}
    \bb{s}_{\omega,\bb{q}} = {i\hbar} \int\!\frac{d\ep}{2\pi\hbar}\int\!\frac{d^2\bb{p}}{(2\pi\hbar)^2}\tr \left[\delta G^<({\ep,\omega,\bb{p},\bb{q}},T) \bb{\sigma}\right],
\end{equation}
where,
\begin{multline}
    \delta G^<(\ep,\omega,\bb{p},\bb{q}) = 1/2 (\delta G^\text{K}(\ep,\omega,\bb{p},\bb{q})\\-
    \delta G^\text{R}(\ep,\omega,\bb{p},\bb{q})+\delta G^\text{A}(\ep,\omega,\bb{p},\bb{q}).  
\end{multline}
In equilibrium we have the fluctuation-dissipation theorem $   G^\text{K}_{\ep_\pm,\bb{p}_\pm} = (1-2f_{\ep_\pm})(G^\text{R}_{\ep_\pm,\bb{p}_\pm}-G^\text{A}_{\ep_\pm,\bb{p}_\pm}) $ with $f_{\ep_\pm}$ the Fermi distribution, so that the spin density now becomes
\begin{align}
    \bb{s}_{\omega,\bb{q}} &= {i\hbar} \int\!\frac{d\ep}{2\pi\hbar}\int\!\frac{d^2\bb{p}}{(2\pi\hbar)^2}\tr\langle\nonumber\\
    & -(f_{\ep_+}-f_{\ep_-})\bb{\sigma}G^\text{R}_{\ep_+,\bb{p}_+} V_{\omega,\bb{q}}G^\text{A}_{\ep_-,\bb{p}_-}\nonumber\\
    &-f_{\ep_+}\bb{\sigma}G^\text{R}_{\ep_+,\bb{p}_+} V_{\omega,\bb{q}}G^\text{R}_{\ep_-,\bb{p}_-}\nonumber\\
    &+f_{\ep_-}\bb{\sigma}G^\text{A}_{\ep_+,\bb{p}_+} V_{\omega,\bb{q}}G^\text{A}_{\ep_-,\bb{p}_-}\rangle,
\end{align}
where the angular brackets stands for impurity averaging. The latter amounts to the replacement of the Green's functions with the corresponding impurity averaged Greens functions (in Born approximation) and to the replacement of one of the spin operators with the corresponding vertex corrected operator (in the non-crossing approximation). The corrections beyond the non-crossing approximation are important for those tensor components that lack leading-order contribution \cite{ivanEPL}. To keep our notations more compact we ignore here the fact that the Green's functions before disorder averaging lack translational invariance, i.\,e. depend on both Wigner coordinates: momentum and coordinate. 

In the limit of small frequency, i.e. $\hbar\omega\ll\ep$, we obtain $s_\alpha = s^\text{I}_\alpha+s^\text{II}_\alpha$,
\begin{align}
    s^\text{I}_\alpha&=\frac{i\omega}{2\hbar}\int\frac{d\ep}{2\pi}\int \frac{d^2\bb{p}}{(2\pi)^2}\Big(-\frac{\partial f}{\partial \ep}\Big)\times\nonumber\\
    &\tr\big\langle
    2\sigma_\alpha G^\text{R}_{\ep_+,\bb{p}_+} V_{\omega,\bb{q}} G^\text{A}_{\ep_-,\bb{p}_-}-\sigma_\alpha G^\text{A}_{\ep_+,\bb{p}_+} V_{\omega,\bb{q}} G^\text{A}_{\ep_-,\bb{p}_-} \nonumber\\
    &\hspace{3.2cm}-\sigma_\alpha G^\text{R}_{\ep_+,\bb{p}_+} V_{\omega,\bb{q}} G^\text{R}_{\ep_-,\bb{p}_-}\big\rangle,\\
    s^\text{II}_\alpha&=\frac{i}{\hbar}\int\frac{d\ep}{2\pi}\int \frac{d^2\bb{p}}{(2\pi)^2} f_\ep\times\nonumber\\
     &\tr \big\langle\sigma_\alpha
    G^\text{A}_{\ep_+,\bb{p}_+} V_{\omega,\bb{q}} G^\text{A}_{\ep_-,\bb{p}_-}-
    \sigma_\alpha G^\text{R}_{\ep_+,\bb{p}_+} V_{\omega,\bb{q}} G^\text{R}_{\ep_-,\bb{p}_-}\big\rangle,
\end{align}
where $s^{\text{I}}$ and $s^\text{II}$ are the Kubo and Streda contributions respectively. The Streda contribution is sub-leading in the powers of weak disorder strength $\alpha\ll 1$ as long as the Fermi energy lies outside the gap. Similarly, the AA and RR bubbles in the expression of $s_\alpha^\text{I}$ are sub-leading and may be neglected. Furthermore, we work in the zero temperature limit. 

The linear response to electric field and time derivative of magnetization corresponds to $V_{\bb{q},\omega} = -\hat{\bb{j}}
\cdot\bb{A}-\Delta_\text{sd} \bb{m}\cdot\bb{\sigma}$, so that we obtain
\begin{equation}
  \bb{s}_{\bb{q},\omega}=\frac{1}{v^2h}\hat{K}(\bb{q},\omega)[ev(\bb{E}_{\bb{q},\omega}\times\hat{\bb{z}})-i\omega\Delta_{\text{sd}}\bb{m}_{\omega}],
  \label{eq:skubo}
\end{equation}
where the components of the tensor $\hat{K}$ are given by
\begin{equation}
\hat{K}_{\alpha\beta}(\bb{q},\omega)={v^2}\int\!\frac{d^2\bb{p}}{(2\pi)^2}\tr \langle \sigma_\alpha G^\text{R}_{\bb{p}+ \hbar q,\ep+\hbar\omega}\sigma_\beta G^\text{A}_{\bb{p},\ep}\rangle.
\label{eq:skuboB}
\end{equation}
Eqs.~(\ref{eq:skubo},\ref{eq:skuboB}) correspond to Eqs.~(\ref{spol},\ref{Kubo}) of the main text. Here we used the expression for the current operator $\hat{\bb{j}} = v_f(\bb{\sigma}\times\hat{\bb{z}})$ and electric field $\bb{E}_{\bb{q},\omega}=i\omega \bb{A}_{\bb{q},\omega}$.

\section{Calculation of the spin-spin correlator} 
\label{app:spinspin}

We shall compute the matrix $\hat{M}$ to the second order in powers of $\omega$ and  $q$. The result is represented as 
\begin{widetext}
\beml
\label{eq:M}
\begin{align}
M &=M_0+M_\omega+M_{\omega^2}+M_{q\omega}+M_{q^2},\\
M_0 &= 
\frac{1}{\pi\alpha(\ep^2+\Delta_z^2)} 
\bpm
\ep^2& 0 & 0 & -\ep \Delta_z \\
 0 & (\ep ^2-\Delta_z^2)/2 & \pi\alpha \ep \Delta_z & 0 \\
 0 & -\pi \alpha \ep \Delta_z & (\ep ^2-\Delta_z^2)/2 & 0 \\
-\ep \Delta_z & 0 & 0 & \Delta_z^2 
\epm,\\
M_\omega & =\frac{i\omega \ep}{\lt[\pi  \alpha  (\ep^2+\Delta_z^2)\rt]^2} 
\bpm
\ep^2 & 0 & 0 &-\ep\Delta_z\\
0 & (\ep ^2-\Delta_z^2)/2 & \pi \alpha (\ep^2-\Delta_z^2)\Delta_z/2\ep& 0 \\
 0 &- \pi \alpha (\ep^2-\Delta_z^2)\Delta_z/2\ep& (\ep ^2-\Delta_z^2)/2  & 0 \\
-\ep\Delta_z& 0 & 0 & \Delta_z^2  
\epm,\\
M_{\omega^2} & =\frac{(i\omega \ep)^2}{\lt[\pi  \alpha  (\ep^2+\Delta_z^2)\rt]^3}
\bpm
\ep^2 & 0 & 0 &-\ep\Delta_z\\
0 & (\ep ^2-\Delta_z^2)/2 & \pi \alpha (\ep^2-\Delta_z^2)\Delta_z/2\ep& 0 \\
 0 & -\pi \alpha (\ep^2-\Delta_z^2)\Delta_z/2\ep& (\ep ^2-\Delta_z^2)/2  & 0 \\
-\ep\Delta_z& 0 & 0 & \Delta_z^2  
\epm,\\
M_{q\omega} & =\frac{v(\ep^2-\Delta_z^2)}{\lt[\pi \alpha \left(\ep^2+\Delta_z^2\right)\rt]^2}\left(\frac{-i}{2}+ \frac{\ep\omega}{\lt[\pi\alpha(\ep^2+\Delta_z^2)\rt]}\right) 
\bpm
 0 & \ep  q_x& \ep  q_y& 0 \\
 \ep  q_x & 0 & 0 & -\Delta_z q_x\\
 \ep  q_y & 0 & 0 & -\Delta_z q_y\\
 0 & -\Delta_z q_x & -\Delta_z q_y & 0 
\epm,\\
M_{q^2} & =\frac{v^2 (\ep^2-\Delta_z^2)}{2\lt[\pi\alpha(\ep^2+\Delta_z^2)\rt]^3} 
\bpm
\ep^2q^2& 0 & 0 & -\ep\Delta_z q^2 \\
 0 & -(\ep^2-\Delta_z^2) (3q_x^2-q_y^2)/4 & - (\ep^2-\Delta_z^2) q_xq_y/2 & 0 \\
 0 & - (\ep^2-\Delta_z^2) q_xq_y/2 & -(\ep^2-\Delta_z^2) (3q_y^2-q_x^2)/4 & 0 \\
-\ep \Delta_z q^2  & 0 & 0 & \Delta_z^2q^2
\epm,
\end{align}
\eml
\end{widetext}
from which the components of $\hat{K}$ are obtained. 
Complete expressions for the components are cumbersome, therefore we proceed by first analyzing their denominator, which is proportional to $\det [1-\pi\alpha M]$
\begin{widetext}
\begin{multline}
  \det [1-\pi\alpha M]=
   -\frac{\varepsilon  \left(\varepsilon ^2+3\Delta_z^2\right)^2}{4 \pi  \alpha  \left(\varepsilon ^2+\Delta_z^2\right)^3}\times\left(i\omega\left(1-i\omega\tau_\text{tr}\frac{\varepsilon^2-5\Delta_z^2}{\varepsilon^2-\Delta_z^2}+\mathcal{O}((\omega\tau_\text{tr})^2)\right)\right.
   \\
   \left.-Dq^2\left(1+i\omega\tau_\text{tr}\frac{13\Delta_z^4+10\Delta_z^2\varepsilon^2+\varepsilon^4}{(\varepsilon^2-\Delta_z^2)(\varepsilon^2+\Delta_z^2)}-(i\omega\tau_\text{tr})^2\frac{(\varepsilon^2+3\Delta^2)(\varepsilon^4-14\varepsilon^2\Delta_z-35\Delta_z^4)}{(\varepsilon^2-\Delta_z^2)(\varepsilon^2+\Delta_z^2)}+\mathcal{O}((\omega\tau_\text{tr})^3)\right)+\mathcal{O}((Dq^2)^2\tau_\text{tr})\right).
\label{eq:sdet}
\end{multline}
\end{widetext}
By restricting ourselves to perturbations that vary slow in time compared to the transport time $\tau_\text{tr}$ and smooth in space compared to the diffusion length $L_D=\sqrt{D\tau_\text{tr}}$, i.e. $\omega\tau_\text{tr},Dq^2\tau_\text{tr}\ll1$, we are able to extract the diffusion pole $(i\omega-Dq^2)^{-1}$.

The components of the conductivity tensor $\hat{\sigma}$ at finite $\omega$ and $\bb{q}$ are given by
\beml
\label{eq:sconductivity}
\begin{align}
  \sigma_{xx} &= \sigma_0+\frac{Dq^2}{i\omega-Dq^2}\left(\frac{q_y^2}{q^2} \sigma_0\right.\nonumber\\
  & \left.-i\omega \tau_\text{tr}\left(\frac{2}{\pi\alpha}\frac{\varepsilon^2+2\Delta_z^2}{\varepsilon^2+\Delta_z^2}+\frac{3}{\pi\alpha}\frac{q_{x}^2-q_{y}^2}{2q^2}\right)\right)\\
  \sigma_{yy} &=  \sigma_0+\frac{Dq^2}{i\omega-Dq^2}\left(\frac{q_x^2}{q^2} \sigma_0\right.\nonumber\\
    &\left.-i\omega \tau_\text{tr}\left(\frac{2}{\pi\alpha}\frac{\varepsilon^2+2\Delta_z^2}{\varepsilon^2+\Delta_z^2}-\frac{3}{\pi\alpha}\frac{q_{x}^2-q_{y}^2}{2q^2}\right)\right)\\
    \sigma_{xy} &= \sigma_\textrm{H}+\frac{Dq^2}{i\omega-Dq^2}\left(-\frac{q_xq_y}{q^2} \sigma_0-i\omega\tau_\text{tr}\frac{3}{\pi\alpha}\frac{q_xq_y}{q^2}\right)\\
  \sigma_{yx} &= -\sigma_\textrm{H}+\frac{Dq^2}{i\omega-Dq^2}\left(-\frac{q_xq_y}{q^2} \sigma_0-i\omega\tau_\text{tr}\frac{3}{\pi\alpha}\frac{q_xq_y}{q^2}\right),
\end{align}
\eml
where $\sigma_0$ and $\sigma_\textrm{H}$ are given in Eq.~(\ref{cond}) of the main text. 
The remaining components of $\hat{K}$ are given by
\beml
\label{Qresult}
\begin{align}
&\bb{Q}=\frac{\Delta_z}{ v}\, \frac{i D \bb{q}}{i\omega-Dq^2}
\lt(1+i\omega\tau_\text{tr}\frac{(\varepsilon^2+7\Delta_z^2)}{\varepsilon^2+\Delta_z^2}\rt),\\
&\zeta=\frac{\Delta_z^2}{ \ep}\,\frac{1-i\omega\tau_\text{tr}(\varepsilon^2-5\Delta_z^2)/(\varepsilon^2-\Delta_z^2)}{i\omega-Dq^2+\omega^2\tau_\text{tr}(\varepsilon^2-5\Delta_z^2)/(\varepsilon^2-\Delta_z^2)}
,
\end{align}
\eml
 where the $\omega^2$-term was included in the denominator of $\zeta$ because of its importance when taking the limit $q\rightarrow 0$.
The leading contributions to Eq.~(\ref{Qresult}a) in the limit $\omega\tau_\text{tr}\ll 1$ together with Eq.~(\ref{Qresult}b) in the limit $q\rightarrow0$ corresponds to Eqs.~(\ref{Kten},\ref{cond},\ref{szz}) of the main text. 

It is convenient to rotate the coordinate system such that the new $\hat{\bb{x}}$ axis lies along $\bb{q}$. Let us introduce a rotation matrix $U$ to transform the tensor $\hat{K}$,  
\begin{equation}
    U = \begin{pmatrix}
            q_x/q & -q_y/q & 0 \\
            q_y/q & q_x/q &0 \\
            0 & 0 & 1
        \end{pmatrix},\hspace{1cm}
    \tilde{K} = U^\top \hat{K} U,
\end{equation}
so that the new components of Eqs.~(\ref{eq:sconductivity}) become
\beml
\label{eq:sconductivity2}
\begin{align}
  \tilde{\sigma}_{xx} &= \sigma_0-\frac{Dq^2}{i\omega-Dq^2}i\omega \tau_\text{tr}\frac{7\varepsilon^2+11\Delta_z^2}{2\pi\alpha (\varepsilon^2+\Delta_z^2)}\\
  \tilde{\sigma}_{yy} &= \sigma_0+\frac{Dq^2}{i\omega-Dq^2} \left(\sigma_0-i\omega \tau_\text{tr}\frac{\varepsilon^2+5\Delta_z^2}{2\pi\alpha(\varepsilon^2+\Delta_z^2)}\right)\\
    \tilde{\sigma}_{yx} &= -\tilde{\sigma}_{xy}=\sigma_\textrm{H},
\end{align}
\eml
and the rotated, $\tilde{K}$, tensor is conveniently written as
\begin{equation}
    \tilde{K} = \begin{pmatrix} \tilde{\sigma}_{xx}         & \sigma_\textrm{H} & 0 \\
                                -\sigma_\textrm{H}  & \tilde{\sigma}_{yy}       & Q \\
                                0                   & Q                 & \zeta
                \end{pmatrix}.
\end{equation}

\section{Limiting behavior of m(t)}
%

To illustrate the behavior of $\bb{m}(t)$ we consider $f=f_0\,\cos(\omega t)$ at a particular point $\bb{r}$. It is, also, convenient to let the field $\bb{H}_\mathrm{eff}$ to lie in the $\hat{\bb{x}}-\hat{\bb{z}}$-plane and rotate the coordinate system such that $\bb{H}_\mathrm{eff}$ lies along new z-direction. This is achieved by introducing the rotation matrix $\hat{R}$,  
\begin{align}
    \hat{R} = \begin{pmatrix} \cos\chi & 0 & -\sin \chi \\ 0 & 1 & 0 \\ \sin\chi & 0 & \cos\chi\end{pmatrix},
\end{align}
where $\chi$ is the angle between $\hat{\bb{z}}$ and $\bb{H}_\mathrm{eff}$. Furthermore, introducing the frequency $\omega_0 = |\gamma\bb{H}_\mathrm{eff}|$ and the unit vector $\bar{\bb{h}}=(-\sin\chi,0,\cos\chi)^\top$, we can write the equation of motion in the rotated coordinate frame as
\begin{multline}
    \partial_t \bb{m} = -\omega_0\,\bb{m}\times \hat{\bb{z}}+f(\bb{r},t)\,\lt(\bb{m}\cdot\bar{\bb{h}}\rt)\;\lt[\bb{m}\times\bar{\bb{h}}\rt]
    \\+
    \alpha_\textrm{G}\, \lt[\bb{m}\times (\partial_t\bb{m})\rt]
    -\alpha_\textrm{G}\, (\partial_t \bb{m})\cdot \hat{\bb{z}}\;\lt[\bb{m}\times \hat{\bb{z}}\rt],
    \label{dynamicEQ}
\end{multline}
where the vector $\hat{\bb{z}}$ is defined now as the unit vector along $\bb{H}_\mathrm{eff}$, hence the magnetization projection $m_\text{H}=\bb{m}\cdot\bb{h}$ is simply given by $m_z$. 



In the regime of $\alpha_\textrm{G}\ll f_0 \ll \ll \omega_0$ we can find the asymptotic behavior of $m_\textrm{H}$ at sufficiently small times. In order to do that it is convenient to represnt $\bb{m}$ in spherical coordinates: $\bb{m}=(\sin\theta\cos\phi,\sin\theta\sin\phi,\cos\theta)^\top$, where $\theta$ is the polar angle between $\bb{m}$ and $\hat{\bb{z}}$ and $\phi$ is the azimuth. In the limit $\alpha_\textrm{G}\rightarrow 0$ we find the equations of motion on $\theta$ and $\phi$:
\begin{align}
\partial_t \theta &= \sin \chi  \sin \phi  f(\bb{r},t) \big(\sin \chi  \sin \theta \cos \phi -\cos \chi  \cos \theta \big)\\
\partial_t \phi &=\omega_0+f(r,t) 
  \cos\theta\big(\cos ^2\chi  \cos ^2\phi-\sin^2\phi \nonumber\\
  &-\frac{1}{2}\sin\chi(\cot\theta-\sin\theta) \big).
\end{align}
We take $f(\bb{r},t)=f_0\cos\omega t$ and assume that $f_0\ll\omega_0$, so that we find $\phi = \omega_0 t-\phi_0$. It is convenient to choose $\phi_0=\pi/2$ so that
\begin{align}
  \partial_t \theta = -f_0 \sin \chi  \cos ^2\omega_0t\,(\sin \theta  \sin \chi  \sin \omega_0t-\cos \theta  \cos \chi ).
  \label{eq:theta2}
\end{align}
Because we assumed that $f_0\ll\omega_0$, the dynamics of $\phi$ is much faster than the dynamics of $\theta$. Therefore we average Eq.~(\ref{eq:theta2}) over $\phi$ and obtain
\begin{equation}
  \partial_t\theta = \frac{f_0}{4} \cos \theta \sin2\chi.
\end{equation}
This equation is readily solved by means of the substitution $\cos\theta = 1/\cosh x$, $\sin\theta=-\tanh x$. Using the initial condition $\theta(0)=0$ one finds 
\begin{equation}
\cos\theta(t) = \frac{1}{\cosh\lt( \frac{1}{4}f_0 t \sin2\chi\rt)},
\end{equation}
which gives the result of Eq.~(\ref{asymptot}) of the main text.


\begin{thebibliography}{69}%
\makeatletter
\providecommand \@ifxundefined [1]{%
 \@ifx{#1\undefined}
}%
\providecommand \@ifnum [1]{%
 \ifnum #1\expandafter \@firstoftwo
 \else \expandafter \@secondoftwo
 \fi
}%
\providecommand \@ifx [1]{%
 \ifx #1\expandafter \@firstoftwo
 \else \expandafter \@secondoftwo
 \fi
}%
\providecommand \natexlab [1]{#1}%
\providecommand \enquote  [1]{``#1''}%
\providecommand \bibnamefont  [1]{#1}%
\providecommand \bibfnamefont [1]{#1}%
\providecommand \citenamefont [1]{#1}%
\providecommand \href@noop [0]{\@secondoftwo}%
\providecommand \href [0]{\begingroup \@sanitize@url \@href}%
\providecommand \@href[1]{\@@startlink{#1}\@@href}%
\providecommand \@@href[1]{\endgroup#1\@@endlink}%
\providecommand \@sanitize@url [0]{\catcode `\\12\catcode `\$12\catcode
  `\&12\catcode `\#12\catcode `\^12\catcode `\_12\catcode `\%12\relax}%
\providecommand \@@startlink[1]{}%
\providecommand \@@endlink[0]{}%
\providecommand \url  [0]{\begingroup\@sanitize@url \@url }%
\providecommand \@url [1]{\endgroup\@href {#1}{\urlprefix }}%
\providecommand \urlprefix  [0]{URL }%
\providecommand \Eprint [0]{\href }%
\providecommand \doibase [0]{http://dx.doi.org/}%
\providecommand \selectlanguage [0]{\@gobble}%
\providecommand \bibinfo  [0]{\@secondoftwo}%
\providecommand \bibfield  [0]{\@secondoftwo}%
\providecommand \translation [1]{[#1]}%
\providecommand \BibitemOpen [0]{}%
\providecommand \bibitemStop [0]{}%
\providecommand \bibitemNoStop [0]{.\EOS\space}%
\providecommand \EOS [0]{\spacefactor3000\relax}%
\providecommand \BibitemShut  [1]{\csname bibitem#1\endcsname}%
\let\auto@bib@innerbib\@empty
\bibitem [{\citenamefont {Miron}\ \emph {et~al.}(2010)\citenamefont {Miron},
  \citenamefont {Gaudin}, \citenamefont {Auffret}, \citenamefont {Rodmacq},
  \citenamefont {Schuhl}, \citenamefont {Pizzini}, \citenamefont {Vogel},\ and\
  \citenamefont {Gambardella}}]{miron_current-driven_2010}%
  \BibitemOpen
  \bibfield  {author} {\bibinfo {author} {\bibfnamefont {I.~M.}\ \bibnamefont
  {Miron}}, \bibinfo {author} {\bibfnamefont {G.}~\bibnamefont {Gaudin}},
  \bibinfo {author} {\bibfnamefont {S.}~\bibnamefont {Auffret}}, \bibinfo
  {author} {\bibfnamefont {B.}~\bibnamefont {Rodmacq}}, \bibinfo {author}
  {\bibfnamefont {A.}~\bibnamefont {Schuhl}}, \bibinfo {author} {\bibfnamefont
  {S.}~\bibnamefont {Pizzini}}, \bibinfo {author} {\bibfnamefont
  {J.}~\bibnamefont {Vogel}}, \ and\ \bibinfo {author} {\bibfnamefont
  {P.}~\bibnamefont {Gambardella}},\ }\href {\doibase 10.1038/nmat2613}
  {\bibfield  {journal} {\bibinfo  {journal} {Nature Materials}\ }\textbf
  {\bibinfo {volume} {9}},\ \bibinfo {pages} {230} (\bibinfo {year}
  {2010})}\BibitemShut {NoStop}%
\bibitem [{\citenamefont {Haney}\ \emph {et~al.}(2013)\citenamefont {Haney},
  \citenamefont {Lee}, \citenamefont {Lee}, \citenamefont {Manchon},\ and\
  \citenamefont {Stiles}}]{haney_current_2013}%
  \BibitemOpen
  \bibfield  {author} {\bibinfo {author} {\bibfnamefont {P.~M.}\ \bibnamefont
  {Haney}}, \bibinfo {author} {\bibfnamefont {H.-W.}\ \bibnamefont {Lee}},
  \bibinfo {author} {\bibfnamefont {K.-J.}\ \bibnamefont {Lee}}, \bibinfo
  {author} {\bibfnamefont {A.}~\bibnamefont {Manchon}}, \ and\ \bibinfo
  {author} {\bibfnamefont {M.~D.}\ \bibnamefont {Stiles}},\ }\href {\doibase
  10.1103/PhysRevB.87.174411} {\bibfield  {journal} {\bibinfo  {journal} {Phys.
  Rev. B}\ }\textbf {\bibinfo {volume} {87}} (\bibinfo {year} {2013}),\
  10.1103/PhysRevB.87.174411}\BibitemShut {NoStop}%
\bibitem [{\citenamefont {Dyakonov}\ and\ \citenamefont
  {Perel}(1971)}]{dyakonov_current-induced_1971}%
  \BibitemOpen
  \bibfield  {author} {\bibinfo {author} {\bibfnamefont {M.~I.}\ \bibnamefont
  {Dyakonov}}\ and\ \bibinfo {author} {\bibfnamefont {V.~I.}\ \bibnamefont
  {Perel}},\ }\href {\doibase 10.1016/0375-9601(71)90196-4} {\bibfield
  {journal} {\bibinfo  {journal} {Physics Letters A}\ }\textbf {\bibinfo
  {volume} {35}},\ \bibinfo {pages} {459} (\bibinfo {year} {1971})}\BibitemShut
  {NoStop}%
\bibitem [{\citenamefont {Awschalom}\ and\ \citenamefont
  {Samarth}(2009)}]{awschalom2009trend}%
  \BibitemOpen
  \bibfield  {author} {\bibinfo {author} {\bibfnamefont {D.}~\bibnamefont
  {Awschalom}}\ and\ \bibinfo {author} {\bibfnamefont {N.}~\bibnamefont
  {Samarth}},\ }\href@noop {} {\bibfield  {journal} {\bibinfo  {journal}
  {Physics}\ }\textbf {\bibinfo {volume} {2}},\ \bibinfo {pages} {50} (\bibinfo
  {year} {2009})}\BibitemShut {NoStop}%
\bibitem [{\citenamefont {Manchon}\ and\ \citenamefont
  {Zhang}(2008)}]{manchon_theory_2008}%
  \BibitemOpen
  \bibfield  {author} {\bibinfo {author} {\bibfnamefont {A.}~\bibnamefont
  {Manchon}}\ and\ \bibinfo {author} {\bibfnamefont {S.}~\bibnamefont
  {Zhang}},\ }\href {\doibase 10.1103/PhysRevB.78.212405} {\bibfield  {journal}
  {\bibinfo  {journal} {Phys. Rev. B}\ }\textbf {\bibinfo {volume} {78}}
  (\bibinfo {year} {2008}),\ 10.1103/PhysRevB.78.212405}\BibitemShut {NoStop}%
\bibitem [{\citenamefont {Garate}\ and\ \citenamefont
  {MacDonald}(2009)}]{garate_influence_2009}%
  \BibitemOpen
  \bibfield  {author} {\bibinfo {author} {\bibfnamefont {I.}~\bibnamefont
  {Garate}}\ and\ \bibinfo {author} {\bibfnamefont {A.~H.}\ \bibnamefont
  {MacDonald}},\ }\href {\doibase 10.1103/PhysRevB.80.134403} {\bibfield
  {journal} {\bibinfo  {journal} {Phys. Rev. B}\ }\textbf {\bibinfo {volume}
  {80}} (\bibinfo {year} {2009}),\ 10.1103/PhysRevB.80.134403}\BibitemShut
  {NoStop}%
\bibitem [{\citenamefont {Manchon}\ and\ \citenamefont
  {Zhang}(2009)}]{manchon2009theory}%
  \BibitemOpen
  \bibfield  {author} {\bibinfo {author} {\bibfnamefont {A.}~\bibnamefont
  {Manchon}}\ and\ \bibinfo {author} {\bibfnamefont {S.}~\bibnamefont
  {Zhang}},\ }\href@noop {} {\bibfield  {journal} {\bibinfo  {journal}
  {Physical Review B}\ }\textbf {\bibinfo {volume} {79}},\ \bibinfo {pages}
  {094422} (\bibinfo {year} {2009})}\BibitemShut {NoStop}%
\bibitem [{\citenamefont
  {Slonczewski}(1996)}]{slonczewski_current-driven_1996}%
  \BibitemOpen
  \bibfield  {author} {\bibinfo {author} {\bibfnamefont {J.~C.}\ \bibnamefont
  {Slonczewski}},\ }\href
  {http://www.sciencedirect.com/science/article/pii/0304885396000625}
  {\bibfield  {journal} {\bibinfo  {journal} {J. Magn. Magn. Mater.}\ }\textbf
  {\bibinfo {volume} {159}},\ \bibinfo {pages} {L1} (\bibinfo {year}
  {1996})}\BibitemShut {NoStop}%
\bibitem [{\citenamefont {Berger}(1996)}]{berger_emission_1996}%
  \BibitemOpen
  \bibfield  {author} {\bibinfo {author} {\bibfnamefont {L.}~\bibnamefont
  {Berger}},\ }\href
  {https://journals.aps.org/prb/abstract/10.1103/PhysRevB.54.9353} {\bibfield
  {journal} {\bibinfo  {journal} {Phys. Rev. B}\ }\textbf {\bibinfo {volume}
  {54}},\ \bibinfo {pages} {9353} (\bibinfo {year} {1996})}\BibitemShut
  {NoStop}%
\bibitem [{\citenamefont {Ralph}\ and\ \citenamefont
  {Stiles}(2008)}]{ralph_spin_2008}%
  \BibitemOpen
  \bibfield  {author} {\bibinfo {author} {\bibfnamefont {D.}~\bibnamefont
  {Ralph}}\ and\ \bibinfo {author} {\bibfnamefont {M.}~\bibnamefont {Stiles}},\
  }\href {\doibase 10.1016/j.jmmm.2007.12.019} {\bibfield  {journal} {\bibinfo
  {journal} {Journal of Magnetism and Magnetic Materials}\ }\textbf {\bibinfo
  {volume} {320}},\ \bibinfo {pages} {1190} (\bibinfo {year}
  {2008})}\BibitemShut {NoStop}%
\bibitem [{\citenamefont {Stiles}\ and\ \citenamefont
  {Zangwill}(2002)}]{stiles_anatomy_2002}%
  \BibitemOpen
  \bibfield  {author} {\bibinfo {author} {\bibfnamefont {M.~D.}\ \bibnamefont
  {Stiles}}\ and\ \bibinfo {author} {\bibfnamefont {A.}~\bibnamefont
  {Zangwill}},\ }\href {\doibase 10.1103/PhysRevB.66.014407} {\bibfield
  {journal} {\bibinfo  {journal} {Physical Review B}\ }\textbf {\bibinfo
  {volume} {66}},\ \bibinfo {pages} {014407} (\bibinfo {year}
  {2002})}\BibitemShut {NoStop}%
\bibitem [{\citenamefont {Fu}\ \emph {et~al.}(2007)\citenamefont {Fu},
  \citenamefont {Kane},\ and\ \citenamefont {Mele}}]{fu_topological_2007}%
  \BibitemOpen
  \bibfield  {author} {\bibinfo {author} {\bibfnamefont {L.}~\bibnamefont
  {Fu}}, \bibinfo {author} {\bibfnamefont {C.~L.}\ \bibnamefont {Kane}}, \ and\
  \bibinfo {author} {\bibfnamefont {E.~J.}\ \bibnamefont {Mele}},\ }\href
  {\doibase 10.1103/PhysRevLett.98.106803} {\bibfield  {journal} {\bibinfo
  {journal} {Physical Review Letters}\ }\textbf {\bibinfo {volume} {98}},\
  \bibinfo {pages} {106803} (\bibinfo {year} {2007})}\BibitemShut {NoStop}%
\bibitem [{\citenamefont {Moore}\ and\ \citenamefont
  {Balents}(2007)}]{moore_topological_2007}%
  \BibitemOpen
  \bibfield  {author} {\bibinfo {author} {\bibfnamefont {J.~E.}\ \bibnamefont
  {Moore}}\ and\ \bibinfo {author} {\bibfnamefont {L.}~\bibnamefont
  {Balents}},\ }\href {\doibase 10.1103/PhysRevB.75.121306} {\bibfield
  {journal} {\bibinfo  {journal} {Physical Review B}\ }\textbf {\bibinfo
  {volume} {75}},\ \bibinfo {pages} {121306} (\bibinfo {year}
  {2007})}\BibitemShut {NoStop}%
\bibitem [{\citenamefont {Roy}(2009)}]{roy_topological_2009}%
  \BibitemOpen
  \bibfield  {author} {\bibinfo {author} {\bibfnamefont {R.}~\bibnamefont
  {Roy}},\ }\href {\doibase 10.1103/PhysRevB.79.195322} {\bibfield  {journal}
  {\bibinfo  {journal} {Physical Review B}\ }\textbf {\bibinfo {volume} {79}},\
  \bibinfo {pages} {195322} (\bibinfo {year} {2009})}\BibitemShut {NoStop}%
\bibitem [{\citenamefont {Hsieh}\ \emph {et~al.}(2008)\citenamefont {Hsieh},
  \citenamefont {Qian}, \citenamefont {Wray}, \citenamefont {Xia},
  \citenamefont {Hor}, \citenamefont {Cava},\ and\ \citenamefont
  {Hasan}}]{hsieh_topological_2008}%
  \BibitemOpen
  \bibfield  {author} {\bibinfo {author} {\bibfnamefont {D.}~\bibnamefont
  {Hsieh}}, \bibinfo {author} {\bibfnamefont {D.}~\bibnamefont {Qian}},
  \bibinfo {author} {\bibfnamefont {L.}~\bibnamefont {Wray}}, \bibinfo {author}
  {\bibfnamefont {Y.}~\bibnamefont {Xia}}, \bibinfo {author} {\bibfnamefont
  {Y.~S.}\ \bibnamefont {Hor}}, \bibinfo {author} {\bibfnamefont {R.~J.}\
  \bibnamefont {Cava}}, \ and\ \bibinfo {author} {\bibfnamefont {M.~Z.}\
  \bibnamefont {Hasan}},\ }\href {\doibase 10.1038/nature06843} {\bibfield
  {journal} {\bibinfo  {journal} {Nature}\ }\textbf {\bibinfo {volume} {452}},\
  \bibinfo {pages} {970} (\bibinfo {year} {2008})}\BibitemShut {NoStop}%
\bibitem [{\citenamefont {Qi}\ \emph {et~al.}(2008{\natexlab{a}})\citenamefont
  {Qi}, \citenamefont {Hughes},\ and\ \citenamefont
  {Zhang}}]{qi_fractional_2008}%
  \BibitemOpen
  \bibfield  {author} {\bibinfo {author} {\bibfnamefont {X.-L.}\ \bibnamefont
  {Qi}}, \bibinfo {author} {\bibfnamefont {T.~L.}\ \bibnamefont {Hughes}}, \
  and\ \bibinfo {author} {\bibfnamefont {S.-C.}\ \bibnamefont {Zhang}},\ }\href
  {\doibase 10.1038/nphys913} {\bibfield  {journal} {\bibinfo  {journal}
  {Nature Physics}\ }\textbf {\bibinfo {volume} {4}},\ \bibinfo {pages} {273}
  (\bibinfo {year} {2008}{\natexlab{a}})}\BibitemShut {NoStop}%
\bibitem [{\citenamefont {Mellnik}\ \emph {et~al.}(2014)\citenamefont
  {Mellnik}, \citenamefont {Lee}, \citenamefont {Richardella}, \citenamefont
  {Grab}, \citenamefont {Mintun}, \citenamefont {Fischer}, \citenamefont
  {Vaezi}, \citenamefont {Manchon}, \citenamefont {Kim}, \citenamefont
  {Samarth},\ and\ \citenamefont {Ralph}}]{mellnik_spin-transfer_2014}%
  \BibitemOpen
  \bibfield  {author} {\bibinfo {author} {\bibfnamefont {A.~R.}\ \bibnamefont
  {Mellnik}}, \bibinfo {author} {\bibfnamefont {J.~S.}\ \bibnamefont {Lee}},
  \bibinfo {author} {\bibfnamefont {A.}~\bibnamefont {Richardella}}, \bibinfo
  {author} {\bibfnamefont {J.~L.}\ \bibnamefont {Grab}}, \bibinfo {author}
  {\bibfnamefont {P.~J.}\ \bibnamefont {Mintun}}, \bibinfo {author}
  {\bibfnamefont {M.~H.}\ \bibnamefont {Fischer}}, \bibinfo {author}
  {\bibfnamefont {A.}~\bibnamefont {Vaezi}}, \bibinfo {author} {\bibfnamefont
  {A.}~\bibnamefont {Manchon}}, \bibinfo {author} {\bibfnamefont {E.-A.}\
  \bibnamefont {Kim}}, \bibinfo {author} {\bibfnamefont {N.}~\bibnamefont
  {Samarth}}, \ and\ \bibinfo {author} {\bibfnamefont {D.~C.}\ \bibnamefont
  {Ralph}},\ }\href {\doibase 10.1038/nature13534} {\bibfield  {journal}
  {\bibinfo  {journal} {Nature}\ }\textbf {\bibinfo {volume} {511}},\ \bibinfo
  {pages} {449} (\bibinfo {year} {2014})}\BibitemShut {NoStop}%
\bibitem [{\citenamefont {Wang}\ \emph {et~al.}(2015)\citenamefont {Wang},
  \citenamefont {Deorani}, \citenamefont {Banerjee}, \citenamefont {Koirala},
  \citenamefont {Brahlek}, \citenamefont {Oh},\ and\ \citenamefont
  {Yang}}]{wang_SOT_BiSe_2014}%
  \BibitemOpen
  \bibfield  {author} {\bibinfo {author} {\bibfnamefont {Y.}~\bibnamefont
  {Wang}}, \bibinfo {author} {\bibfnamefont {P.}~\bibnamefont {Deorani}},
  \bibinfo {author} {\bibfnamefont {K.}~\bibnamefont {Banerjee}}, \bibinfo
  {author} {\bibfnamefont {N.}~\bibnamefont {Koirala}}, \bibinfo {author}
  {\bibfnamefont {M.}~\bibnamefont {Brahlek}}, \bibinfo {author} {\bibfnamefont
  {S.}~\bibnamefont {Oh}}, \ and\ \bibinfo {author} {\bibfnamefont
  {H.}~\bibnamefont {Yang}},\ }\href {\doibase 10.1103/PhysRevLett.114.257202}
  {\bibfield  {journal} {\bibinfo  {journal} {Phys. Rev. Lett.}\ }\textbf
  {\bibinfo {volume} {114}},\ \bibinfo {pages} {257202} (\bibinfo {year}
  {2015})}\BibitemShut {NoStop}%
\bibitem [{\citenamefont {Fan}\ \emph {et~al.}(2014)\citenamefont {Fan},
  \citenamefont {Upadhyaya}, \citenamefont {Kou}, \citenamefont {Lang},
  \citenamefont {Takei}, \citenamefont {Wang}, \citenamefont {Tang},
  \citenamefont {He}, \citenamefont {Chang}, \citenamefont {Montazeri},
  \citenamefont {Yu}, \citenamefont {Jiang}, \citenamefont {Nie}, \citenamefont
  {Schwartz}, \citenamefont {Tserkovnyak},\ and\ \citenamefont
  {Wang}}]{fan_magnetization_2014}%
  \BibitemOpen
  \bibfield  {author} {\bibinfo {author} {\bibfnamefont {Y.}~\bibnamefont
  {Fan}}, \bibinfo {author} {\bibfnamefont {P.}~\bibnamefont {Upadhyaya}},
  \bibinfo {author} {\bibfnamefont {X.}~\bibnamefont {Kou}}, \bibinfo {author}
  {\bibfnamefont {M.}~\bibnamefont {Lang}}, \bibinfo {author} {\bibfnamefont
  {S.}~\bibnamefont {Takei}}, \bibinfo {author} {\bibfnamefont
  {Z.}~\bibnamefont {Wang}}, \bibinfo {author} {\bibfnamefont {J.}~\bibnamefont
  {Tang}}, \bibinfo {author} {\bibfnamefont {L.}~\bibnamefont {He}}, \bibinfo
  {author} {\bibfnamefont {L.-T.}\ \bibnamefont {Chang}}, \bibinfo {author}
  {\bibfnamefont {M.}~\bibnamefont {Montazeri}}, \bibinfo {author}
  {\bibfnamefont {G.}~\bibnamefont {Yu}}, \bibinfo {author} {\bibfnamefont
  {W.}~\bibnamefont {Jiang}}, \bibinfo {author} {\bibfnamefont
  {T.}~\bibnamefont {Nie}}, \bibinfo {author} {\bibfnamefont {R.~N.}\
  \bibnamefont {Schwartz}}, \bibinfo {author} {\bibfnamefont {Y.}~\bibnamefont
  {Tserkovnyak}}, \ and\ \bibinfo {author} {\bibfnamefont {K.~L.}\ \bibnamefont
  {Wang}},\ }\href {\doibase 10.1038/nmat3973} {\bibfield  {journal} {\bibinfo
  {journal} {Nature Materials}\ }\textbf {\bibinfo {volume} {13}},\ \bibinfo
  {pages} {699} (\bibinfo {year} {2014})}\BibitemShut {NoStop}%
\bibitem [{\citenamefont {Fan}\ \emph {et~al.}(2016)\citenamefont {Fan},
  \citenamefont {Kou}, \citenamefont {Upadhyaya}, \citenamefont {Shao},
  \citenamefont {Pan}, \citenamefont {Lang}, \citenamefont {Che}, \citenamefont
  {Tang}, \citenamefont {Montazeri}, \citenamefont {Murata}, \citenamefont
  {Chang}, \citenamefont {Akyol}, \citenamefont {Yu}, \citenamefont {Nie},
  \citenamefont {Wong}, \citenamefont {Liu}, \citenamefont {Wang},
  \citenamefont {Tserkovnyak},\ and\ \citenamefont {Wang}}]{Fan_SOT_TI_2016}%
  \BibitemOpen
  \bibfield  {author} {\bibinfo {author} {\bibfnamefont {Y.}~\bibnamefont
  {Fan}}, \bibinfo {author} {\bibfnamefont {X.}~\bibnamefont {Kou}}, \bibinfo
  {author} {\bibfnamefont {P.}~\bibnamefont {Upadhyaya}}, \bibinfo {author}
  {\bibfnamefont {Q.}~\bibnamefont {Shao}}, \bibinfo {author} {\bibfnamefont
  {L.}~\bibnamefont {Pan}}, \bibinfo {author} {\bibfnamefont {M.}~\bibnamefont
  {Lang}}, \bibinfo {author} {\bibfnamefont {X.}~\bibnamefont {Che}}, \bibinfo
  {author} {\bibfnamefont {J.}~\bibnamefont {Tang}}, \bibinfo {author}
  {\bibfnamefont {M.}~\bibnamefont {Montazeri}}, \bibinfo {author}
  {\bibfnamefont {K.}~\bibnamefont {Murata}}, \bibinfo {author} {\bibfnamefont
  {L.-T.}\ \bibnamefont {Chang}}, \bibinfo {author} {\bibfnamefont
  {M.}~\bibnamefont {Akyol}}, \bibinfo {author} {\bibfnamefont
  {G.}~\bibnamefont {Yu}}, \bibinfo {author} {\bibfnamefont {T.}~\bibnamefont
  {Nie}}, \bibinfo {author} {\bibfnamefont {K.~L.}\ \bibnamefont {Wong}},
  \bibinfo {author} {\bibfnamefont {J.}~\bibnamefont {Liu}}, \bibinfo {author}
  {\bibfnamefont {Y.}~\bibnamefont {Wang}}, \bibinfo {author} {\bibfnamefont
  {Y.}~\bibnamefont {Tserkovnyak}}, \ and\ \bibinfo {author} {\bibfnamefont
  {K.~L.}\ \bibnamefont {Wang}},\ }\href
  {http://dx.doi.org/10.1038/nnano.2015.294} {\bibfield  {journal} {\bibinfo
  {journal} {Nature Nanotechnology}\ }\textbf {\bibinfo {volume} {11}},\
  \bibinfo {pages} {352 EP } (\bibinfo {year} {2016})}\BibitemShut {NoStop}%
\bibitem [{\citenamefont {Yasuda}\ \emph {et~al.}(2017)\citenamefont {Yasuda},
  \citenamefont {Tsukazaki}, \citenamefont {Yoshimi}, \citenamefont {Kondou},
  \citenamefont {Takahashi}, \citenamefont {Otani}, \citenamefont {Kawasaki},\
  and\ \citenamefont {Tokura}}]{Yasuda_SOT_BiSbTe_2017}%
  \BibitemOpen
  \bibfield  {author} {\bibinfo {author} {\bibfnamefont {K.}~\bibnamefont
  {Yasuda}}, \bibinfo {author} {\bibfnamefont {A.}~\bibnamefont {Tsukazaki}},
  \bibinfo {author} {\bibfnamefont {R.}~\bibnamefont {Yoshimi}}, \bibinfo
  {author} {\bibfnamefont {K.}~\bibnamefont {Kondou}}, \bibinfo {author}
  {\bibfnamefont {K.~S.}\ \bibnamefont {Takahashi}}, \bibinfo {author}
  {\bibfnamefont {Y.}~\bibnamefont {Otani}}, \bibinfo {author} {\bibfnamefont
  {M.}~\bibnamefont {Kawasaki}}, \ and\ \bibinfo {author} {\bibfnamefont
  {Y.}~\bibnamefont {Tokura}},\ }\href {\doibase
  10.1103/PhysRevLett.119.137204} {\bibfield  {journal} {\bibinfo  {journal}
  {Phys. Rev. Lett.}\ }\textbf {\bibinfo {volume} {119}},\ \bibinfo {pages}
  {137204} (\bibinfo {year} {2017})}\BibitemShut {NoStop}%
\bibitem [{\citenamefont {Cha}\ \emph {et~al.}(2018)\citenamefont {Cha},
  \citenamefont {Noh}, \citenamefont {Kim}, \citenamefont {Son}, \citenamefont
  {Bae}, \citenamefont {Lee}, \citenamefont {Kim}, \citenamefont {Lee},
  \citenamefont {Shin}, \citenamefont {Sim}, \citenamefont {Yang},
  \citenamefont {Lee}, \citenamefont {Shim}, \citenamefont {Lee}, \citenamefont
  {Jo}, \citenamefont {Kim}, \citenamefont {Kim},\ and\ \citenamefont
  {Choi}}]{Cha2018}%
  \BibitemOpen
  \bibfield  {author} {\bibinfo {author} {\bibfnamefont {S.}~\bibnamefont
  {Cha}}, \bibinfo {author} {\bibfnamefont {M.}~\bibnamefont {Noh}}, \bibinfo
  {author} {\bibfnamefont {J.}~\bibnamefont {Kim}}, \bibinfo {author}
  {\bibfnamefont {J.}~\bibnamefont {Son}}, \bibinfo {author} {\bibfnamefont
  {H.}~\bibnamefont {Bae}}, \bibinfo {author} {\bibfnamefont {D.}~\bibnamefont
  {Lee}}, \bibinfo {author} {\bibfnamefont {H.}~\bibnamefont {Kim}}, \bibinfo
  {author} {\bibfnamefont {J.}~\bibnamefont {Lee}}, \bibinfo {author}
  {\bibfnamefont {H.-S.}\ \bibnamefont {Shin}}, \bibinfo {author}
  {\bibfnamefont {S.}~\bibnamefont {Sim}}, \bibinfo {author} {\bibfnamefont
  {S.}~\bibnamefont {Yang}}, \bibinfo {author} {\bibfnamefont {S.}~\bibnamefont
  {Lee}}, \bibinfo {author} {\bibfnamefont {W.}~\bibnamefont {Shim}}, \bibinfo
  {author} {\bibfnamefont {C.-H.}\ \bibnamefont {Lee}}, \bibinfo {author}
  {\bibfnamefont {M.-H.}\ \bibnamefont {Jo}}, \bibinfo {author} {\bibfnamefont
  {J.~S.}\ \bibnamefont {Kim}}, \bibinfo {author} {\bibfnamefont
  {D.}~\bibnamefont {Kim}}, \ and\ \bibinfo {author} {\bibfnamefont
  {H.}~\bibnamefont {Choi}},\ }\href {\doibase 10.1038/s41565-018-0195-y}
  {\bibfield  {journal} {\bibinfo  {journal} {Nature Nanotechnology}\ }
  (\bibinfo {year} {2018}),\ 10.1038/s41565-018-0195-y}\BibitemShut {NoStop}%
\bibitem [{\citenamefont {Liu}\ \emph {et~al.}(2011)\citenamefont {Liu},
  \citenamefont {Moriyama}, \citenamefont {Ralph},\ and\ \citenamefont
  {Buhrman}}]{Ralph2011SOTFMresonance}%
  \BibitemOpen
  \bibfield  {author} {\bibinfo {author} {\bibfnamefont {L.}~\bibnamefont
  {Liu}}, \bibinfo {author} {\bibfnamefont {T.}~\bibnamefont {Moriyama}},
  \bibinfo {author} {\bibfnamefont {D.~C.}\ \bibnamefont {Ralph}}, \ and\
  \bibinfo {author} {\bibfnamefont {R.~A.}\ \bibnamefont {Buhrman}},\ }\href
  {\doibase 10.1103/PhysRevLett.106.036601} {\bibfield  {journal} {\bibinfo
  {journal} {Phys. Rev. Lett.}\ }\textbf {\bibinfo {volume} {106}},\ \bibinfo
  {pages} {036601} (\bibinfo {year} {2011})}\BibitemShut {NoStop}%
\bibitem [{\citenamefont {Akyol}\ \emph {et~al.}(2015)\citenamefont {Akyol},
  \citenamefont {Yu}, \citenamefont {Alzate}, \citenamefont {Upadhyaya},
  \citenamefont {Li}, \citenamefont {Wong}, \citenamefont {Ekicibil},
  \citenamefont {Khalili~Amiri},\ and\ \citenamefont
  {Wang}}]{Wang2015_SOTHflCoFEBIMgO}%
  \BibitemOpen
  \bibfield  {author} {\bibinfo {author} {\bibfnamefont {M.}~\bibnamefont
  {Akyol}}, \bibinfo {author} {\bibfnamefont {G.}~\bibnamefont {Yu}}, \bibinfo
  {author} {\bibfnamefont {J.~G.}\ \bibnamefont {Alzate}}, \bibinfo {author}
  {\bibfnamefont {P.}~\bibnamefont {Upadhyaya}}, \bibinfo {author}
  {\bibfnamefont {X.}~\bibnamefont {Li}}, \bibinfo {author} {\bibfnamefont
  {K.~L.}\ \bibnamefont {Wong}}, \bibinfo {author} {\bibfnamefont
  {A.}~\bibnamefont {Ekicibil}}, \bibinfo {author} {\bibfnamefont
  {P.}~\bibnamefont {Khalili~Amiri}}, \ and\ \bibinfo {author} {\bibfnamefont
  {K.~L.}\ \bibnamefont {Wang}},\ }\href {\doibase 10.1063/1.4919108}
  {\bibfield  {journal} {\bibinfo  {journal} {Applied Physics Letters}\
  }\textbf {\bibinfo {volume} {106}},\ \bibinfo {pages} {162409} (\bibinfo
  {year} {2015})},\ \Eprint
  {http://arxiv.org/abs/https://doi.org/10.1063/1.4919108}
  {https://doi.org/10.1063/1.4919108} \BibitemShut {NoStop}%
\bibitem [{\citenamefont {MacNeill}\ \emph {et~al.}(2016)\citenamefont
  {MacNeill}, \citenamefont {Stiehl}, \citenamefont {Guimaraes}, \citenamefont
  {Buhrman}, \citenamefont {Park},\ and\ \citenamefont
  {Ralph}}]{Ralph2016SOTWTE2}%
  \BibitemOpen
  \bibfield  {author} {\bibinfo {author} {\bibfnamefont {D.}~\bibnamefont
  {MacNeill}}, \bibinfo {author} {\bibfnamefont {G.~M.}\ \bibnamefont
  {Stiehl}}, \bibinfo {author} {\bibfnamefont {M.~H.~D.}\ \bibnamefont
  {Guimaraes}}, \bibinfo {author} {\bibfnamefont {R.~A.}\ \bibnamefont
  {Buhrman}}, \bibinfo {author} {\bibfnamefont {J.}~\bibnamefont {Park}}, \
  and\ \bibinfo {author} {\bibfnamefont {D.~C.}\ \bibnamefont {Ralph}},\ }\href
  {https://doi.org/10.1038/nphys3933} {\bibfield  {journal} {\bibinfo
  {journal} {Nature Physics}\ }\textbf {\bibinfo {volume} {13}},\ \bibinfo
  {pages} {300 EP } (\bibinfo {year} {2016})}\BibitemShut {NoStop}%
\bibitem [{\citenamefont {Guimaraes}\ \emph {et~al.}(2018)\citenamefont
  {Guimaraes}, \citenamefont {Stiehl}, \citenamefont {MacNeill}, \citenamefont
  {Reynolds},\ and\ \citenamefont {Ralph}}]{Ralph2018SOT}%
  \BibitemOpen
  \bibfield  {author} {\bibinfo {author} {\bibfnamefont {M.~H.~D.}\
  \bibnamefont {Guimaraes}}, \bibinfo {author} {\bibfnamefont {G.~M.}\
  \bibnamefont {Stiehl}}, \bibinfo {author} {\bibfnamefont {D.}~\bibnamefont
  {MacNeill}}, \bibinfo {author} {\bibfnamefont {N.~D.}\ \bibnamefont
  {Reynolds}}, \ and\ \bibinfo {author} {\bibfnamefont {D.~C.}\ \bibnamefont
  {Ralph}},\ }\href {\doibase 10.1021/acs.nanolett.7b04993} {\bibfield
  {journal} {\bibinfo  {journal} {Nano Letters}\ }\textbf {\bibinfo {volume}
  {18}},\ \bibinfo {pages} {1311} (\bibinfo {year} {2018})},\ \bibinfo {note}
  {pMID: 29328662},\ \Eprint
  {http://arxiv.org/abs/https://doi.org/10.1021/acs.nanolett.7b04993}
  {https://doi.org/10.1021/acs.nanolett.7b04993} \BibitemShut {NoStop}%
\bibitem [{\citenamefont {Sakai}\ and\ \citenamefont
  {Kohno}(2014)}]{sakai_spin_2014}%
  \BibitemOpen
  \bibfield  {author} {\bibinfo {author} {\bibfnamefont {A.}~\bibnamefont
  {Sakai}}\ and\ \bibinfo {author} {\bibfnamefont {H.}~\bibnamefont {Kohno}},\
  }\href {\doibase 10.1103/PhysRevB.89.165307} {\bibfield  {journal} {\bibinfo
  {journal} {Physical Review B}\ }\textbf {\bibinfo {volume} {89}} (\bibinfo
  {year} {2014}),\ 10.1103/PhysRevB.89.165307}\BibitemShut {NoStop}%
\bibitem [{\citenamefont {Ndiaye}\ \emph {et~al.}(2017)\citenamefont {Ndiaye},
  \citenamefont {Akosa}, \citenamefont {Fischer}, \citenamefont {Vaezi},
  \citenamefont {Kim},\ and\ \citenamefont {Manchon}}]{ndiaye_dirac_2017}%
  \BibitemOpen
  \bibfield  {author} {\bibinfo {author} {\bibfnamefont {P.~B.}\ \bibnamefont
  {Ndiaye}}, \bibinfo {author} {\bibfnamefont {C.~A.}\ \bibnamefont {Akosa}},
  \bibinfo {author} {\bibfnamefont {M.~H.}\ \bibnamefont {Fischer}}, \bibinfo
  {author} {\bibfnamefont {A.}~\bibnamefont {Vaezi}}, \bibinfo {author}
  {\bibfnamefont {E.-A.}\ \bibnamefont {Kim}}, \ and\ \bibinfo {author}
  {\bibfnamefont {A.}~\bibnamefont {Manchon}},\ }\href {\doibase
  10.1103/PhysRevB.96.014408} {\bibfield  {journal} {\bibinfo  {journal}
  {Physical Review B}\ }\textbf {\bibinfo {volume} {96}} (\bibinfo {year}
  {2017}),\ 10.1103/PhysRevB.96.014408}\BibitemShut {NoStop}%
\bibitem [{\citenamefont {Burkov}\ and\ \citenamefont
  {Hawthorn}(2010)}]{burkov_spin_2010}%
  \BibitemOpen
  \bibfield  {author} {\bibinfo {author} {\bibfnamefont {A.~A.}\ \bibnamefont
  {Burkov}}\ and\ \bibinfo {author} {\bibfnamefont {D.~G.}\ \bibnamefont
  {Hawthorn}},\ }\href {\doibase 10.1103/PhysRevLett.105.066802} {\bibfield
  {journal} {\bibinfo  {journal} {Physical Review Letters}\ }\textbf {\bibinfo
  {volume} {105}},\ \bibinfo {pages} {066802} (\bibinfo {year}
  {2010})}\BibitemShut {NoStop}%
\bibitem [{\citenamefont {Taguchi}\ \emph {et~al.}(2015)\citenamefont
  {Taguchi}, \citenamefont {Shintani},\ and\ \citenamefont
  {Tanaka}}]{taguchi_spin-charge_2015}%
  \BibitemOpen
  \bibfield  {author} {\bibinfo {author} {\bibfnamefont {K.}~\bibnamefont
  {Taguchi}}, \bibinfo {author} {\bibfnamefont {K.}~\bibnamefont {Shintani}}, \
  and\ \bibinfo {author} {\bibfnamefont {Y.}~\bibnamefont {Tanaka}},\ }\href
  {\doibase 10.1103/PhysRevB.92.035425} {\bibfield  {journal} {\bibinfo
  {journal} {Physical Review B}\ }\textbf {\bibinfo {volume} {92}} (\bibinfo
  {year} {2015}),\ 10.1103/PhysRevB.92.035425}\BibitemShut {NoStop}%
\bibitem [{\citenamefont {Shintani}\ \emph {et~al.}(2016)\citenamefont
  {Shintani}, \citenamefont {Taguchi}, \citenamefont {Tanaka},\ and\
  \citenamefont {Kawaguchi}}]{shintani_spin_2016}%
  \BibitemOpen
  \bibfield  {author} {\bibinfo {author} {\bibfnamefont {K.}~\bibnamefont
  {Shintani}}, \bibinfo {author} {\bibfnamefont {K.}~\bibnamefont {Taguchi}},
  \bibinfo {author} {\bibfnamefont {Y.}~\bibnamefont {Tanaka}}, \ and\ \bibinfo
  {author} {\bibfnamefont {Y.}~\bibnamefont {Kawaguchi}},\ }\href {\doibase
  10.1103/PhysRevB.93.195415} {\bibfield  {journal} {\bibinfo  {journal}
  {Physical Review B}\ }\textbf {\bibinfo {volume} {93}},\ \bibinfo {pages}
  {195415} (\bibinfo {year} {2016})}\BibitemShut {NoStop}%
\bibitem [{\citenamefont {Vonsovsky}(1974)}]{sdmodel}%
  \BibitemOpen
  \bibfield  {author} {\bibinfo {author} {\bibfnamefont {S.~V.}\ \bibnamefont
  {Vonsovsky}},\ }\href@noop {} {\emph {\bibinfo {title} {Magnetism}}}\
  (\bibinfo  {publisher} {Wiley, New York},\ \bibinfo {year}
  {1974})\BibitemShut {NoStop}%
\bibitem [{\citenamefont {Ghosh}\ and\ \citenamefont
  {Manchon}(2018)}]{Ghosh18}%
  \BibitemOpen
  \bibfield  {author} {\bibinfo {author} {\bibfnamefont {S.}~\bibnamefont
  {Ghosh}}\ and\ \bibinfo {author} {\bibfnamefont {A.}~\bibnamefont
  {Manchon}},\ }\href {\doibase 10.1103/PhysRevB.97.134402} {\bibfield
  {journal} {\bibinfo  {journal} {Phys. Rev. B}\ }\textbf {\bibinfo {volume}
  {97}},\ \bibinfo {pages} {134402} (\bibinfo {year} {2018})}\BibitemShut
  {NoStop}%
\bibitem [{\citenamefont {Maleev}\ and\ \citenamefont
  {Toperverg}(1975)}]{maleev1975corrections}%
  \BibitemOpen
  \bibfield  {author} {\bibinfo {author} {\bibfnamefont {S.}~\bibnamefont
  {Maleev}}\ and\ \bibinfo {author} {\bibfnamefont {B.}~\bibnamefont
  {Toperverg}},\ }\href@noop {} {\bibfield  {journal} {\bibinfo  {journal} {Zh.
  Eksp. Teor. Fiz}\ }\textbf {\bibinfo {volume} {69}},\ \bibinfo {pages} {1440}
  (\bibinfo {year} {1975})}\BibitemShut {NoStop}%
\bibitem [{\citenamefont {Maleev}\ and\ \citenamefont
  {Toperverg}(1976)}]{maleev1976corrections}%
  \BibitemOpen
  \bibfield  {author} {\bibinfo {author} {\bibfnamefont {S.}~\bibnamefont
  {Maleev}}\ and\ \bibinfo {author} {\bibfnamefont {B.}~\bibnamefont
  {Toperverg}},\ }\href@noop {} {\bibfield  {journal} {\bibinfo  {journal}
  {Soviet Journal of Experimental and Theoretical Physics}\ }\textbf {\bibinfo
  {volume} {42}},\ \bibinfo {pages} {734} (\bibinfo {year} {1976})}\BibitemShut
  {NoStop}%
\bibitem [{\citenamefont {Altshuler}\ and\ \citenamefont
  {Aronov}(1985)}]{altshuler1985electron}%
  \BibitemOpen
  \bibfield  {author} {\bibinfo {author} {\bibfnamefont {B.}~\bibnamefont
  {Altshuler}}\ and\ \bibinfo {author} {\bibfnamefont {A.~G.}\ \bibnamefont
  {Aronov}},\ }\href@noop {} {\bibfield  {journal} {\bibinfo  {journal}
  {Amsterdam: North-Holland) p}\ }\textbf {\bibinfo {volume} {1}},\ \bibinfo
  {pages} {155} (\bibinfo {year} {1985})}\BibitemShut {NoStop}%
\bibitem [{\citenamefont {Ado}\ \emph {et~al.}(2015)\citenamefont {Ado},
  \citenamefont {Dmitriev}, \citenamefont {Ostrovsky},\ and\ \citenamefont
  {Titov}}]{ivanEPL}%
  \BibitemOpen
  \bibfield  {author} {\bibinfo {author} {\bibfnamefont {I.~A.}\ \bibnamefont
  {Ado}}, \bibinfo {author} {\bibfnamefont {I.~A.}\ \bibnamefont {Dmitriev}},
  \bibinfo {author} {\bibfnamefont {P.~M.}\ \bibnamefont {Ostrovsky}}, \ and\
  \bibinfo {author} {\bibfnamefont {M.}~\bibnamefont {Titov}},\ }\href
  {\doibase 10.1209/0295-5075/111/37004} {\bibfield  {journal} {\bibinfo
  {journal} {EPL (Europhysics Letters)}\ }\textbf {\bibinfo {volume} {111}},\
  \bibinfo {pages} {37004} (\bibinfo {year} {2015})}\BibitemShut {NoStop}%
\bibitem [{\citenamefont {Ado}\ \emph {et~al.}(2016)\citenamefont {Ado},
  \citenamefont {Dmitriev}, \citenamefont {Ostrovsky},\ and\ \citenamefont
  {Titov}}]{ivanPRL}%
  \BibitemOpen
  \bibfield  {author} {\bibinfo {author} {\bibfnamefont {I.~A.}\ \bibnamefont
  {Ado}}, \bibinfo {author} {\bibfnamefont {I.~A.}\ \bibnamefont {Dmitriev}},
  \bibinfo {author} {\bibfnamefont {P.~M.}\ \bibnamefont {Ostrovsky}}, \ and\
  \bibinfo {author} {\bibfnamefont {M.}~\bibnamefont {Titov}},\ }\href
  {\doibase 10.1103/PhysRevLett.117.046601} {\bibfield  {journal} {\bibinfo
  {journal} {Phys. Rev. Lett.}\ }\textbf {\bibinfo {volume} {117}},\ \bibinfo
  {pages} {046601} (\bibinfo {year} {2016})}\BibitemShut {NoStop}%
\bibitem [{\citenamefont {Ado}\ \emph {et~al.}(2017)\citenamefont {Ado},
  \citenamefont {Dmitriev}, \citenamefont {Ostrovsky},\ and\ \citenamefont
  {Titov}}]{ivanPRB}%
  \BibitemOpen
  \bibfield  {author} {\bibinfo {author} {\bibfnamefont {I.~A.}\ \bibnamefont
  {Ado}}, \bibinfo {author} {\bibfnamefont {I.~A.}\ \bibnamefont {Dmitriev}},
  \bibinfo {author} {\bibfnamefont {P.~M.}\ \bibnamefont {Ostrovsky}}, \ and\
  \bibinfo {author} {\bibfnamefont {M.}~\bibnamefont {Titov}},\ }\href
  {\doibase 10.1103/PhysRevB.96.235148} {\bibfield  {journal} {\bibinfo
  {journal} {Phys. Rev. B}\ }\textbf {\bibinfo {volume} {96}},\ \bibinfo
  {pages} {235148} (\bibinfo {year} {2017})}\BibitemShut {NoStop}%
\bibitem [{\citenamefont {Kong}\ \emph {et~al.}(2011)\citenamefont {Kong},
  \citenamefont {Cha}, \citenamefont {Lai}, \citenamefont {Peng}, \citenamefont
  {Analytis}, \citenamefont {Meister}, \citenamefont {Chen}, \citenamefont
  {Zhang}, \citenamefont {Fisher}, \citenamefont {Shen},\ and\ \citenamefont
  {Cui}}]{kong_rapid_2011}%
  \BibitemOpen
  \bibfield  {author} {\bibinfo {author} {\bibfnamefont {D.}~\bibnamefont
  {Kong}}, \bibinfo {author} {\bibfnamefont {J.~J.}\ \bibnamefont {Cha}},
  \bibinfo {author} {\bibfnamefont {K.}~\bibnamefont {Lai}}, \bibinfo {author}
  {\bibfnamefont {H.}~\bibnamefont {Peng}}, \bibinfo {author} {\bibfnamefont
  {J.~G.}\ \bibnamefont {Analytis}}, \bibinfo {author} {\bibfnamefont
  {S.}~\bibnamefont {Meister}}, \bibinfo {author} {\bibfnamefont
  {Y.}~\bibnamefont {Chen}}, \bibinfo {author} {\bibfnamefont {H.-J.}\
  \bibnamefont {Zhang}}, \bibinfo {author} {\bibfnamefont {I.~R.}\ \bibnamefont
  {Fisher}}, \bibinfo {author} {\bibfnamefont {Z.-X.}\ \bibnamefont {Shen}}, \
  and\ \bibinfo {author} {\bibfnamefont {Y.}~\bibnamefont {Cui}},\ }\href
  {\doibase 10.1021/nn200556h} {\bibfield  {journal} {\bibinfo  {journal} {ACS
  Nano}\ }\textbf {\bibinfo {volume} {5}},\ \bibinfo {pages} {4698} (\bibinfo
  {year} {2011})}\BibitemShut {NoStop}%
\bibitem [{\citenamefont {Kamboj}\ \emph {et~al.}(2017)\citenamefont {Kamboj},
  \citenamefont {Singh}, \citenamefont {Ferrus}, \citenamefont {Beere},
  \citenamefont {Duffy}, \citenamefont {Hesjedal}, \citenamefont {Barnes},\
  and\ \citenamefont {Ritchie}}]{kamboj_probing_2017}%
  \BibitemOpen
  \bibfield  {author} {\bibinfo {author} {\bibfnamefont {V.~S.}\ \bibnamefont
  {Kamboj}}, \bibinfo {author} {\bibfnamefont {A.}~\bibnamefont {Singh}},
  \bibinfo {author} {\bibfnamefont {T.}~\bibnamefont {Ferrus}}, \bibinfo
  {author} {\bibfnamefont {H.~E.}\ \bibnamefont {Beere}}, \bibinfo {author}
  {\bibfnamefont {L.~B.}\ \bibnamefont {Duffy}}, \bibinfo {author}
  {\bibfnamefont {T.}~\bibnamefont {Hesjedal}}, \bibinfo {author}
  {\bibfnamefont {C.~H.~W.}\ \bibnamefont {Barnes}}, \ and\ \bibinfo {author}
  {\bibfnamefont {D.~A.}\ \bibnamefont {Ritchie}},\ }\href {\doibase
  10.1021/acsphotonics.7b00492} {\bibfield  {journal} {\bibinfo  {journal} {ACS
  Photonics}\ }\textbf {\bibinfo {volume} {4}},\ \bibinfo {pages} {2711}
  (\bibinfo {year} {2017})}\BibitemShut {NoStop}%
\bibitem [{\citenamefont {Huang}\ \emph {et~al.}(2017)\citenamefont {Huang},
  \citenamefont {Huang}, \citenamefont {Hsu}, \citenamefont {Wadekar},
  \citenamefont {Yan}, \citenamefont {Yu},\ and\ \citenamefont
  {Chou}}]{huang_enhancement_2017}%
  \BibitemOpen
  \bibfield  {author} {\bibinfo {author} {\bibfnamefont {S.-M.}\ \bibnamefont
  {Huang}}, \bibinfo {author} {\bibfnamefont {S.-J.}\ \bibnamefont {Huang}},
  \bibinfo {author} {\bibfnamefont {C.}~\bibnamefont {Hsu}}, \bibinfo {author}
  {\bibfnamefont {P.~V.}\ \bibnamefont {Wadekar}}, \bibinfo {author}
  {\bibfnamefont {Y.-J.}\ \bibnamefont {Yan}}, \bibinfo {author} {\bibfnamefont
  {S.-H.}\ \bibnamefont {Yu}}, \ and\ \bibinfo {author} {\bibfnamefont
  {M.}~\bibnamefont {Chou}},\ }\href {\doibase 10.1038/s41598-017-05369-y}
  {\bibfield  {journal} {\bibinfo  {journal} {Scientific Reports}\ }\textbf
  {\bibinfo {volume} {7}} (\bibinfo {year} {2017}),\
  10.1038/s41598-017-05369-y}\BibitemShut {NoStop}%
\bibitem [{\citenamefont {Xiang}\ \emph {et~al.}(2014)\citenamefont {Xiang},
  \citenamefont {Wang},\ and\ \citenamefont {Dou}}]{xiang_transport_2014}%
  \BibitemOpen
  \bibfield  {author} {\bibinfo {author} {\bibfnamefont {F.~X.}\ \bibnamefont
  {Xiang}}, \bibinfo {author} {\bibfnamefont {X.~L.}\ \bibnamefont {Wang}}, \
  and\ \bibinfo {author} {\bibfnamefont {S.~X.}\ \bibnamefont {Dou}},\ }\href
  {http://arxiv.org/abs/1401.6732} {\bibfield  {journal} {\bibinfo  {journal}
  {arXiv:1401.6732 [cond-mat]}\ } (\bibinfo {year} {2014})},\ \bibinfo {note}
  {arXiv: 1401.6732}\BibitemShut {NoStop}%
\bibitem [{\citenamefont {van~der Bijl}\ and\ \citenamefont
  {Duine}(2012)}]{vanderBijl2012}%
  \BibitemOpen
  \bibfield  {author} {\bibinfo {author} {\bibfnamefont {E.}~\bibnamefont
  {van~der Bijl}}\ and\ \bibinfo {author} {\bibfnamefont {R.~A.}\ \bibnamefont
  {Duine}},\ }\href {\doibase 10.1103/PhysRevB.86.094406} {\bibfield  {journal}
  {\bibinfo  {journal} {Phys. Rev. B}\ }\textbf {\bibinfo {volume} {86}},\
  \bibinfo {pages} {094406} (\bibinfo {year} {2012})}\BibitemShut {NoStop}%
\bibitem [{\citenamefont {Hals}\ and\ \citenamefont
  {Brataas}(2013)}]{Hals2013}%
  \BibitemOpen
  \bibfield  {author} {\bibinfo {author} {\bibfnamefont {K.~M.~D.}\
  \bibnamefont {Hals}}\ and\ \bibinfo {author} {\bibfnamefont {A.}~\bibnamefont
  {Brataas}},\ }\href {\doibase 10.1103/PhysRevB.88.085423} {\bibfield
  {journal} {\bibinfo  {journal} {Phys. Rev. B}\ }\textbf {\bibinfo {volume}
  {88}},\ \bibinfo {pages} {085423} (\bibinfo {year} {2013})}\BibitemShut
  {NoStop}%
\bibitem [{\citenamefont {Tserkovnyak}\ and\ \citenamefont
  {Wong}(2009)}]{tserkovnyak_theory_2009}%
  \BibitemOpen
  \bibfield  {author} {\bibinfo {author} {\bibfnamefont {Y.}~\bibnamefont
  {Tserkovnyak}}\ and\ \bibinfo {author} {\bibfnamefont {C.~H.}\ \bibnamefont
  {Wong}},\ }\href {\doibase 10.1103/PhysRevB.79.014402} {\bibfield  {journal}
  {\bibinfo  {journal} {Phys. Rev. B}\ }\textbf {\bibinfo {volume} {79}},\
  \bibinfo {pages} {014402} (\bibinfo {year} {2009})}\BibitemShut {NoStop}%
\bibitem [{\citenamefont {Mahfouzi}\ \emph {et~al.}(2012)\citenamefont
  {Mahfouzi}, \citenamefont {Nagaosa},\ and\ \citenamefont
  {Nikoli{\'c}}}]{mahfouzi_spin-orbit_2012}%
  \BibitemOpen
  \bibfield  {author} {\bibinfo {author} {\bibfnamefont {F.}~\bibnamefont
  {Mahfouzi}}, \bibinfo {author} {\bibfnamefont {N.}~\bibnamefont {Nagaosa}}, \
  and\ \bibinfo {author} {\bibfnamefont {B.~K.}\ \bibnamefont {Nikoli{\'c}}},\
  }\href {\doibase 10.1103/PhysRevLett.109.166602} {\bibfield  {journal}
  {\bibinfo  {journal} {Physical Review Letters}\ }\textbf {\bibinfo {volume}
  {109}} (\bibinfo {year} {2012}),\ 10.1103/PhysRevLett.109.166602}\BibitemShut
  {NoStop}%
\bibitem [{\citenamefont {Ferreiros}\ \emph {et~al.}(2015)\citenamefont
  {Ferreiros}, \citenamefont {Buijnsters},\ and\ \citenamefont
  {Katsnelson}}]{katsnelson15}%
  \BibitemOpen
  \bibfield  {author} {\bibinfo {author} {\bibfnamefont {Y.}~\bibnamefont
  {Ferreiros}}, \bibinfo {author} {\bibfnamefont {F.~J.}\ \bibnamefont
  {Buijnsters}}, \ and\ \bibinfo {author} {\bibfnamefont {M.~I.}\ \bibnamefont
  {Katsnelson}},\ }\href {\doibase 10.1103/PhysRevB.92.085416} {\bibfield
  {journal} {\bibinfo  {journal} {Phys. Rev. B}\ }\textbf {\bibinfo {volume}
  {92}},\ \bibinfo {pages} {085416} (\bibinfo {year} {2015})}\BibitemShut
  {NoStop}%
\bibitem [{\citenamefont {Fischer}\ \emph {et~al.}(2016)\citenamefont
  {Fischer}, \citenamefont {Vaezi}, \citenamefont {Manchon},\ and\
  \citenamefont {Kim}}]{fischer_spin-torque_2016}%
  \BibitemOpen
  \bibfield  {author} {\bibinfo {author} {\bibfnamefont {M.~H.}\ \bibnamefont
  {Fischer}}, \bibinfo {author} {\bibfnamefont {A.}~\bibnamefont {Vaezi}},
  \bibinfo {author} {\bibfnamefont {A.}~\bibnamefont {Manchon}}, \ and\
  \bibinfo {author} {\bibfnamefont {E.-A.}\ \bibnamefont {Kim}},\ }\href
  {\doibase 10.1103/PhysRevB.93.125303} {\bibfield  {journal} {\bibinfo
  {journal} {Physical Review B}\ }\textbf {\bibinfo {volume} {93}} (\bibinfo
  {year} {2016}),\ 10.1103/PhysRevB.93.125303},\ \Eprint
  {http://arxiv.org/abs/1305.1328} {arXiv:1305.1328} \BibitemShut {NoStop}%
\bibitem [{\citenamefont {Yokoyama}\ \emph {et~al.}(2010)\citenamefont
  {Yokoyama}, \citenamefont {Zang},\ and\ \citenamefont
  {Nagaosa}}]{yokoyama_theoretical_2010}%
  \BibitemOpen
  \bibfield  {author} {\bibinfo {author} {\bibfnamefont {T.}~\bibnamefont
  {Yokoyama}}, \bibinfo {author} {\bibfnamefont {J.}~\bibnamefont {Zang}}, \
  and\ \bibinfo {author} {\bibfnamefont {N.}~\bibnamefont {Nagaosa}},\ }\href
  {\doibase 10.1103/PhysRevB.81.241410} {\bibfield  {journal} {\bibinfo
  {journal} {Physical Review B}\ }\textbf {\bibinfo {volume} {81}} (\bibinfo
  {year} {2010}),\ 10.1103/PhysRevB.81.241410}\BibitemShut {NoStop}%
\bibitem [{\citenamefont {Yokoyama}(2011)}]{yokoyama_current-induced_2011}%
  \BibitemOpen
  \bibfield  {author} {\bibinfo {author} {\bibfnamefont {T.}~\bibnamefont
  {Yokoyama}},\ }\href {\doibase 10.1103/PhysRevB.84.113407} {\bibfield
  {journal} {\bibinfo  {journal} {Physical Review B}\ }\textbf {\bibinfo
  {volume} {84}},\ \bibinfo {pages} {113407} (\bibinfo {year}
  {2011})}\BibitemShut {NoStop}%
\bibitem [{\citenamefont {Siu}\ \emph {et~al.}(2016)\citenamefont {Siu},
  \citenamefont {Son}, \citenamefont {Jalil},\ and\ \citenamefont
  {Tan}}]{siu_spin_2016}%
  \BibitemOpen
  \bibfield  {author} {\bibinfo {author} {\bibfnamefont {Z.~B.}\ \bibnamefont
  {Siu}}, \bibinfo {author} {\bibfnamefont {H.~C.}\ \bibnamefont {Son}},
  \bibinfo {author} {\bibfnamefont {M.~b.~A.}\ \bibnamefont {Jalil}}, \ and\
  \bibinfo {author} {\bibfnamefont {S.~G.}\ \bibnamefont {Tan}},\ }\href
  {http://arxiv.org/abs/1609.02242} {\bibfield  {journal} {\bibinfo  {journal}
  {arXiv:1609.02242 [cond-mat]}\ } (\bibinfo {year} {2016})},\ \Eprint
  {http://arxiv.org/abs/1609.02242} {arXiv:1609.02242 [cond-mat]} \BibitemShut
  {NoStop}%
\bibitem [{\citenamefont {Mahfouzi}\ \emph {et~al.}(2016)\citenamefont
  {Mahfouzi}, \citenamefont {Nikoli{\'c}},\ and\ \citenamefont
  {Kioussis}}]{mahfouzi_antidamping_2016}%
  \BibitemOpen
  \bibfield  {author} {\bibinfo {author} {\bibfnamefont {F.}~\bibnamefont
  {Mahfouzi}}, \bibinfo {author} {\bibfnamefont {B.~K.}\ \bibnamefont
  {Nikoli{\'c}}}, \ and\ \bibinfo {author} {\bibfnamefont {N.}~\bibnamefont
  {Kioussis}},\ }\href {\doibase 10.1103/PhysRevB.93.115419} {\bibfield
  {journal} {\bibinfo  {journal} {Physical Review B}\ }\textbf {\bibinfo
  {volume} {93}} (\bibinfo {year} {2016}),\
  10.1103/PhysRevB.93.115419}\BibitemShut {NoStop}%
\bibitem [{\citenamefont {Soleimani}\ \emph {et~al.}(2017)\citenamefont
  {Soleimani}, \citenamefont {Jalili}, \citenamefont {Mahfouzi},\ and\
  \citenamefont {Kioussis}}]{soleimani_spin-orbit_2017}%
  \BibitemOpen
  \bibfield  {author} {\bibinfo {author} {\bibfnamefont {M.}~\bibnamefont
  {Soleimani}}, \bibinfo {author} {\bibfnamefont {S.}~\bibnamefont {Jalili}},
  \bibinfo {author} {\bibfnamefont {F.}~\bibnamefont {Mahfouzi}}, \ and\
  \bibinfo {author} {\bibfnamefont {N.}~\bibnamefont {Kioussis}},\ }\href
  {\doibase 10.1209/0295-5075/117/37001} {\bibfield  {journal} {\bibinfo
  {journal} {EPL (Europhysics Letters)}\ }\textbf {\bibinfo {volume} {117}},\
  \bibinfo {pages} {37001} (\bibinfo {year} {2017})}\BibitemShut {NoStop}%
\bibitem [{\citenamefont {Kurebayashi}\ and\ \citenamefont
  {Nomura}(2017)}]{kurebayashi_microscopic_2017}%
  \BibitemOpen
  \bibfield  {author} {\bibinfo {author} {\bibfnamefont {D.}~\bibnamefont
  {Kurebayashi}}\ and\ \bibinfo {author} {\bibfnamefont {K.}~\bibnamefont
  {Nomura}},\ }\href {http://arxiv.org/abs/1702.04918} {\bibfield  {journal}
  {\bibinfo  {journal} {arXiv:1702.04918 [cond-mat]}\ } (\bibinfo {year}
  {2017})},\ \Eprint {http://arxiv.org/abs/1702.04918} {arXiv:1702.04918
  [cond-mat]} \BibitemShut {NoStop}%
\bibitem [{\citenamefont {Chen}\ \emph {et~al.}(2017)\citenamefont {Chen},
  \citenamefont {Peng},\ and\ \citenamefont
  {Zhou}}]{chen_current-induced_2017}%
  \BibitemOpen
  \bibfield  {author} {\bibinfo {author} {\bibfnamefont {J.}~\bibnamefont
  {Chen}}, \bibinfo {author} {\bibfnamefont {Y.}~\bibnamefont {Peng}}, \ and\
  \bibinfo {author} {\bibfnamefont {J.}~\bibnamefont {Zhou}},\ }\href {\doibase
  10.1016/j.jmmm.2017.02.043} {\bibfield  {journal} {\bibinfo  {journal}
  {Journal of Magnetism and Magnetic Materials}\ }\textbf {\bibinfo {volume}
  {432}},\ \bibinfo {pages} {554} (\bibinfo {year} {2017})}\BibitemShut
  {NoStop}%
\bibitem [{\citenamefont {Rodriguez-Vega}\ \emph {et~al.}(2016)\citenamefont
  {Rodriguez-Vega}, \citenamefont {Schwiete}, \citenamefont {Sinova},\ and\
  \citenamefont {Rossi}}]{rodriguez-vega_giant_2016}%
  \BibitemOpen
  \bibfield  {author} {\bibinfo {author} {\bibfnamefont {M.}~\bibnamefont
  {Rodriguez-Vega}}, \bibinfo {author} {\bibfnamefont {G.}~\bibnamefont
  {Schwiete}}, \bibinfo {author} {\bibfnamefont {J.}~\bibnamefont {Sinova}}, \
  and\ \bibinfo {author} {\bibfnamefont {E.}~\bibnamefont {Rossi}},\ }\href
  {http://arxiv.org/abs/1610.04229} {\bibfield  {journal} {\bibinfo  {journal}
  {arXiv:1610.04229 [cond-mat]}\ } (\bibinfo {year} {2016})},\ \Eprint
  {http://arxiv.org/abs/1610.04229} {arXiv:1610.04229 [cond-mat]} \BibitemShut
  {NoStop}%
\bibitem [{\citenamefont {Qi}\ \emph {et~al.}(2008{\natexlab{b}})\citenamefont
  {Qi}, \citenamefont {Hughes},\ and\ \citenamefont
  {Zhang}}]{qi_topological_2008}%
  \BibitemOpen
  \bibfield  {author} {\bibinfo {author} {\bibfnamefont {X.-L.}\ \bibnamefont
  {Qi}}, \bibinfo {author} {\bibfnamefont {T.~L.}\ \bibnamefont {Hughes}}, \
  and\ \bibinfo {author} {\bibfnamefont {S.-C.}\ \bibnamefont {Zhang}},\ }\href
  {\doibase 10.1103/PhysRevB.78.195424} {\bibfield  {journal} {\bibinfo
  {journal} {Physical Review B}\ }\textbf {\bibinfo {volume} {78}},\ \bibinfo
  {pages} {195424} (\bibinfo {year} {2008}{\natexlab{b}})}\BibitemShut
  {NoStop}%
\bibitem [{\citenamefont {Garate}\ and\ \citenamefont
  {Franz}(2010)}]{garate_inverse_2010}%
  \BibitemOpen
  \bibfield  {author} {\bibinfo {author} {\bibfnamefont {I.}~\bibnamefont
  {Garate}}\ and\ \bibinfo {author} {\bibfnamefont {M.}~\bibnamefont {Franz}},\
  }\href {\doibase 10.1103/PhysRevLett.104.146802} {\bibfield  {journal}
  {\bibinfo  {journal} {Physical Review Letters}\ }\textbf {\bibinfo {volume}
  {104}},\ \bibinfo {pages} {146802} (\bibinfo {year} {2010})}\BibitemShut
  {NoStop}%
\bibitem [{\citenamefont {Nomura}\ and\ \citenamefont
  {Nagaosa}(2010)}]{nomura_electric_2010}%
  \BibitemOpen
  \bibfield  {author} {\bibinfo {author} {\bibfnamefont {K.}~\bibnamefont
  {Nomura}}\ and\ \bibinfo {author} {\bibfnamefont {N.}~\bibnamefont
  {Nagaosa}},\ }\href {\doibase 10.1103/PhysRevB.82.161401} {\bibfield
  {journal} {\bibinfo  {journal} {Physical Review B}\ }\textbf {\bibinfo
  {volume} {82}},\ \bibinfo {pages} {161401} (\bibinfo {year}
  {2010})}\BibitemShut {NoStop}%
\bibitem [{\citenamefont {Tserkovnyak}\ and\ \citenamefont
  {Loss}(2012)}]{tserkovnyak_thin-film_2012-1}%
  \BibitemOpen
  \bibfield  {author} {\bibinfo {author} {\bibfnamefont {Y.}~\bibnamefont
  {Tserkovnyak}}\ and\ \bibinfo {author} {\bibfnamefont {D.}~\bibnamefont
  {Loss}},\ }\href {\doibase 10.1103/PhysRevLett.108.187201} {\bibfield
  {journal} {\bibinfo  {journal} {Physical Review Letters}\ }\textbf {\bibinfo
  {volume} {108}} (\bibinfo {year} {2012}),\
  10.1103/PhysRevLett.108.187201}\BibitemShut {NoStop}%
\bibitem [{\citenamefont {Linder}(2014)}]{linder_improved_2014}%
  \BibitemOpen
  \bibfield  {author} {\bibinfo {author} {\bibfnamefont {J.}~\bibnamefont
  {Linder}},\ }\href {\doibase 10.1103/PhysRevB.90.041412} {\bibfield
  {journal} {\bibinfo  {journal} {Physical Review B}\ }\textbf {\bibinfo
  {volume} {90}},\ \bibinfo {pages} {041412} (\bibinfo {year}
  {2014})}\BibitemShut {NoStop}%
\bibitem [{\citenamefont {Tserkovnyak}\ \emph {et~al.}(2015)\citenamefont
  {Tserkovnyak}, \citenamefont {Pesin},\ and\ \citenamefont
  {Loss}}]{tserkovnyak_spin_2015}%
  \BibitemOpen
  \bibfield  {author} {\bibinfo {author} {\bibfnamefont {Y.}~\bibnamefont
  {Tserkovnyak}}, \bibinfo {author} {\bibfnamefont {D.~A.}\ \bibnamefont
  {Pesin}}, \ and\ \bibinfo {author} {\bibfnamefont {D.}~\bibnamefont {Loss}},\
  }\href {\doibase 10.1103/PhysRevB.91.041121} {\bibfield  {journal} {\bibinfo
  {journal} {Physical Review B}\ }\textbf {\bibinfo {volume} {91}},\ \bibinfo
  {pages} {041121} (\bibinfo {year} {2015})}\BibitemShut {NoStop}%
\bibitem [{\citenamefont {Ueda}\ \emph {et~al.}(2012)\citenamefont {Ueda},
  \citenamefont {Takeuchi}, \citenamefont {Tatara},\ and\ \citenamefont
  {Yokoyama}}]{ueda_topological_2012}%
  \BibitemOpen
  \bibfield  {author} {\bibinfo {author} {\bibfnamefont {H.~T.}\ \bibnamefont
  {Ueda}}, \bibinfo {author} {\bibfnamefont {A.}~\bibnamefont {Takeuchi}},
  \bibinfo {author} {\bibfnamefont {G.}~\bibnamefont {Tatara}}, \ and\ \bibinfo
  {author} {\bibfnamefont {T.}~\bibnamefont {Yokoyama}},\ }\href {\doibase
  10.1103/PhysRevB.85.115110} {\bibfield  {journal} {\bibinfo  {journal}
  {Physical Review B}\ }\textbf {\bibinfo {volume} {85}},\ \bibinfo {pages}
  {115110} (\bibinfo {year} {2012})}\BibitemShut {NoStop}%
\bibitem [{\citenamefont {Liu}\ and\ \citenamefont
  {Sinova}(2013)}]{liu_reading_2013}%
  \BibitemOpen
  \bibfield  {author} {\bibinfo {author} {\bibfnamefont {X.}~\bibnamefont
  {Liu}}\ and\ \bibinfo {author} {\bibfnamefont {J.}~\bibnamefont {Sinova}},\
  }\href {\doibase 10.1103/PhysRevLett.111.166801} {\bibfield  {journal}
  {\bibinfo  {journal} {Physical Review Letters}\ }\textbf {\bibinfo {volume}
  {111}},\ \bibinfo {pages} {166801} (\bibinfo {year} {2013})}\BibitemShut
  {NoStop}%
\bibitem [{\citenamefont {Chang}\ \emph {et~al.}(2015)\citenamefont {Chang},
  \citenamefont {Markussen}, \citenamefont {Smidstrup}, \citenamefont
  {Stokbro},\ and\ \citenamefont {Nikoli{\'c}}}]{chang_nonequilibrium_2015}%
  \BibitemOpen
  \bibfield  {author} {\bibinfo {author} {\bibfnamefont {P.-H.}\ \bibnamefont
  {Chang}}, \bibinfo {author} {\bibfnamefont {T.}~\bibnamefont {Markussen}},
  \bibinfo {author} {\bibfnamefont {S.}~\bibnamefont {Smidstrup}}, \bibinfo
  {author} {\bibfnamefont {K.}~\bibnamefont {Stokbro}}, \ and\ \bibinfo
  {author} {\bibfnamefont {B.~K.}\ \bibnamefont {Nikoli{\'c}}},\ }\href
  {\doibase 10.1103/PhysRevB.92.201406} {\bibfield  {journal} {\bibinfo
  {journal} {Physical Review B}\ }\textbf {\bibinfo {volume} {92}},\ \bibinfo
  {pages} {201406} (\bibinfo {year} {2015})}\BibitemShut {NoStop}%
\bibitem [{\citenamefont {Fujimoto}\ and\ \citenamefont
  {Kohno}(2014)}]{fujimoto_transport_2014}%
  \BibitemOpen
  \bibfield  {author} {\bibinfo {author} {\bibfnamefont {J.}~\bibnamefont
  {Fujimoto}}\ and\ \bibinfo {author} {\bibfnamefont {H.}~\bibnamefont
  {Kohno}},\ }\href {\doibase 10.1103/PhysRevB.90.214418} {\bibfield  {journal}
  {\bibinfo  {journal} {Physical Review B}\ }\textbf {\bibinfo {volume} {90}},\
  \bibinfo {pages} {214418} (\bibinfo {year} {2014})}\BibitemShut {NoStop}%
\bibitem [{\citenamefont {Okuma}\ and\ \citenamefont
  {Ogata}(2016)}]{okuma_unconventional_2016}%
  \BibitemOpen
  \bibfield  {author} {\bibinfo {author} {\bibfnamefont {N.}~\bibnamefont
  {Okuma}}\ and\ \bibinfo {author} {\bibfnamefont {M.}~\bibnamefont {Ogata}},\
  }\href {\doibase 10.1103/PhysRevB.93.140205} {\bibfield  {journal} {\bibinfo
  {journal} {Physical Review B}\ }\textbf {\bibinfo {volume} {93}},\ \bibinfo
  {pages} {140205} (\bibinfo {year} {2016})}\BibitemShut {NoStop}%
\bibitem [{\citenamefont {Rammer}\ and\ \citenamefont
  {Smith}(1986)}]{rammer_quantum_1986}%
  \BibitemOpen
  \bibfield  {author} {\bibinfo {author} {\bibfnamefont {J.}~\bibnamefont
  {Rammer}}\ and\ \bibinfo {author} {\bibfnamefont {H.}~\bibnamefont {Smith}},\
  }\href {http://journals.aps.org/rmp/abstract/10.1103/RevModPhys.58.323}
  {\bibfield  {journal} {\bibinfo  {journal} {Reviews of Modern Physics}\
  }\textbf {\bibinfo {volume} {58}},\ \bibinfo {pages} {323} (\bibinfo {year}
  {1986})}\BibitemShut {NoStop}%
\end{thebibliography}
\end{document}